\documentclass{aa}
\usepackage[varg]{txfonts}
\usepackage{natbib}
\usepackage{url}
\usepackage{hyperref}
\usepackage{xcolor}
\usepackage{ulem}
\usepackage{orcidlink}
\usepackage{xspace}

\usepackage{upgreek}
\usepackage{graphicx}   % Including figure files
\usepackage{multirow}
\usepackage[switch]{linenoaa}   %    \linenumbers 
\newcommand{\fourU}{{4U~1538$-$52}\xspace}
\newcommand{\ixpe}{{IXPE}\xspace}
\newcommand{\msun}{M_{\odot}}  

\begin{document}

\title{Exploring polarization and geometry in the X-ray pulsar \fourU}
\titlerunning{X-ray polarimetry of \fourU}

\authorrunning{Loktev, V., et al.}

\author{Vladislav~Loktev \inst{\ref{in:UTU},\ref{in:Hel}}\orcidlink{0000-0001-6894-871X}
\and Sofia~V.~Forsblom \inst{\ref{in:UTU}}\orcidlink{0000-0001-9167-2790}
\and Sergey~S.~Tsygankov \inst{\ref{in:UTU},\ref{in:IKI}}\orcidlink{0000-0002-9679-0793}
\and Juri~Poutanen \inst{\ref{in:UTU},\ref{in:IKI}}\orcidlink{0000-0002-0983-0049} 
\and Alexander~A.~Mushtukov \inst{\ref{in:Oxford}}\orcidlink{0000-0003-2306-419X} 
%\and Victor~Doroshenko\inst{\ref{in:Tub}}
\and Alessandro~Di~Marco\inst{\ref{in:INAF-IAPS}}\orcidlink{0000-0003-0331-3259}
\and Jeremy~Heyl \inst{\ref{in:UBC}}\orcidlink{0000-0001-9739-367X} 
\and Ruth~M.~E.~Kelly \inst{\ref{in:MSSL},\ref{in:MIT}}\orcidlink{0000-0002-5004-3573}
\and Fabio~La~Monaca  \inst{\ref{in:INAF-IAPS},\ref{in:UniRoma2}}\orcidlink{0000-0001-8916-4156} 
\and Mason~Ng \inst{\ref{in:McGill},\ref{in:McGill-TSI}}\orcidlink{0000-0002-0940-6563} 
\and Swati~Ravi \inst{\ref{in:MIT}}\orcidlink{0000-0002-2381-4184}
\and Alexander~Salganik \inst{\ref{in:UTU}}\orcidlink{0000-0003-2609-8838} 
\and Andrea~Santangelo\inst{\ref{in:Tub}}\orcidlink{0000-0003-4187-9560}
\and Valery~F.~Suleimanov\inst{\ref{in:Tub}}\orcidlink{0000-0003-3733-7267}
\and Silvia~Zane  \inst{\ref{in:MSSL}}\orcidlink{0000-0001-5326-880X}  
}

\institute{
Department of Physics and Astronomy, 20014 University of Turku,  Finland \label{in:UTU}\\
\email{vladislav.loktev@utu.fi}
\and Department of Physics, P.O. Box 64, 00014 University of Helsinki, Finland \label{in:Hel}
\and Space Research Institute, Russian Academy of Sciences, Profsoyuznaya 84/32, Moscow 117997, Russia \label{in:IKI} 
\and Astrophysics, Department of Physics, University of Oxford, Denys Wilkinson Building, Keble Road, Oxford OX1 3RH, UK \label{in:Oxford}
\and  INAF Istituto di Astrofisica e Planetologia Spaziali, Via del Fosso del Cavaliere 100, 00133 Roma, Italy \label{in:INAF-IAPS}
\and University of British Columbia, Vancouver, BC V6T 1Z4, Canada \label{in:UBC}
\and Mullard Space Science Laboratory, University College London, Holmbury St Mary, Dorking, Surrey RH5 6NT, UK \label{in:MSSL}
\and MIT Kavli Institute for Astrophysics and Space Research, Massachusetts Institute of Technology, 77 Massachusetts Avenue, Cambridge, MA 02139, USA \label{in:MIT} 
\and Dipartimento di Fisica, Universit\`{a} degli Studi di Roma ``Tor Vergata'', Via della Ricerca Scientifica 1, 00133 Roma, Italy \label{in:UniRoma2}
\and Department of Physics, McGill University, 3600 rue University, Montr\'{e}al, QC H3A 2T8, Canada \label{in:McGill}
\and Trottier Space Institute, McGill University, 3550 rue University, Montr\'{e}al, QC H3A 2A7, Canada \label{in:McGill-TSI}
\and Institut f\"ur Astronomie und Astrophysik, Universit\"at T\"ubingen, Sand 1, 72076 T\"ubingen, Germany \label{in:Tub}
}

%\date{Received xxx / Accepted xxx}

\abstract
{The Imaging X-ray Polarimetry Explorer (\ixpe) observations of accreting X-ray pulsars (XRPs) continue to provide novel insights into the physics and geometry of these sources. 
We present the first X-ray polarimetric study of the persistent wind-fed XRP \fourU, based on five \ixpe\ observations totaling 360 ks, conducted in March and October 2024.
We detect marginally significant polarization in the combined data set in the full 2--8 keV energy band, with a polarization degree (PD) of $3.0\pm1.1$\% and polarization angle (PA) of $-18\degr\pm11\degr$. 
The energy-resolved analysis shows a clear energy dependence of the polarization properties, with a remarkable $\sim$70\degr\ switch in PA between low and high energies. 
Similarly, the pulse phase-resolved spectro-polarimetric analysis reveals different signatures at low and high energies. At low (2--3~keV) energies, the PD ranges between $\sim$2\% and $\sim$18\%, with the PA varying between $-16\degr$ and 70\degr. At higher (4--8~keV) energies, the PD varies between $\sim$3\% and $\sim$12\%, with a drastically different PA behavior.
Fitting the rotating vector model to the pulse phase dependence of the PA at the lower energies, we constrain the geometric configuration of the pulsar.
The analysis favors a high spin-axis inclination of $>50\degr$, which agrees with both previous pulse-phase-dependent spectral fitting of the cyclotron line region and the known high orbital inclination of the binary system. 
The magnetic obliquity is estimated to be $30\degr$ and the spin position angle to be $19\degr$. 
A sharp switch in PA around 3 keV presents a particular theoretical challenge, as it is not consistent with the right-angle switch that was only seen in one other pulsar, \mbox{Vela~X-1}. 
}

\keywords{magnetic fields -- methods: observational -- polarization -- pulsars:  individual: \fourU\ -- stars: neutron -- X-rays: binaries}

\maketitle

\section{Introduction}

Accreting X-ray pulsars (XRPs) are highly magnetized neutron stars (NSs) in binary systems, where matter is transferred from a massive optical companion via an accretion disk or stellar wind. The presence of a strong magnetic field, typically in the range of $10^{12}$--$10^{13}$~G, significantly affects the accretion process, guiding the inflowing plasma along magnetic field lines toward the NS magnetic poles. This results in the formation of compact emission regions of different structure, hot spots or extended accretion columns, depending on the mass accretion rate \citep[see][for a recent review]{MushtukovTsygankov2024}. 
These systems have been extensively studied with timing and spectroscopy for the past decades, but the geometric configuration of NSs and host binaries cannot be probed directly using these methods.

With the launch of the Imaging X-ray Polarimetry Explorer (\ixpe) in December 2021, X-ray polarimetry in the 2--8~keV band became an indispensable tool to constrain the geometric parameters of XRPs. 
Within the first three years of its operation, \ixpe\ observed a dozen   XRPs: 
\mbox{Her~X-1} \citep{2022NatAs...6.1433D,Heyl2024}, 
 \mbox{Cen~X-3} \citep{Tsygankov22},
 \mbox{X~Persei} \citep{2023MNRAS.524.2004M}, 
 \mbox{4U~1626$-$67} \citep{2022ApJ...940...70M},
 \mbox{Vela~X-1} \citep{Forsblom23},
 \mbox{GRO~J1008$-$57} \citep{2023A&A...675A..48T},
 \mbox{EXO~2030+375} \citep{2023A&A...675A..29M},
 \mbox{LS~V~+44~17} \citep{Doroshenko23},
 \mbox{GX 301$-$2} \citep{2023A&A...678A.119S},
 \mbox{Swift J0243.6$+$6124} \citep{Poutanen2024},
and \mbox{SMC~X-1} \citep{Forsblom24}. 
The observed polarization degree (PD) in all these sources turned out to be drastically different from theoretical expectations, which predicted high polarization of up to 60--80\%  \citep{1988ApJ...324.1056M,2006RPPh...69.2631H,2021MNRAS.501..109C}. In reality, less than 25\% was observed, even in phase-resolved data \citep[for a review see][]{Poutanen2024Galax}.
The observations of low PDs in XRPs have led to new discussions on the physical origin of polarization in these objects. 
The development of updated models requires expanding the sample of XRPs observed in polarized X-rays and accreting under different conditions.

The binary system \fourU\ (sometimes referred to as 4U~1538$-$522) is a persistent XRP, in which a NS accretes from the stellar wind of its companion star, a 20~\,$\msun$~B0.2\,Ia type supergiant, QV~Nor \citep{1990ApJ...353..274C,2023-HMXBcat}. 
Initially discovered with a pulsation period of approximately 529~s \citep{1977ApJ...216L..11B,1977MNRAS.179P..35D}, the pulsar has exhibited long-term spin-up and spin-down trends, which have been documented over the course of several decades \citep{1997ApJ...488..413R,2006A&A...453.1037B,2020ApJ...896...90M,Tamang2024}.
The most recent value of 526.23~s was found using the \textit{NuSTAR} data \citep{Tamang2024}. 
The pulse profile shows a simple two-peak structure in soft X-rays with a weak dependence on luminosity \citep{Tamang2024}.
%The spin axis inclination and magnetic obliquity of the NS in \fourU\ remained impossible to constrain from the pulse profile. 
Certain progress has been made in constraining the geometrical configuration of the polar caps in \fourU\ using X-ray spectral data from \textit{Ginga} observations \citep{Bulik1992}. 
By modeling phase-dependent spectra around the cyclotron line at 20 keV, \citet{Bulik1995} inferred the size, location, and temperature of the polar caps.
The analysis indicated significant differences between them, suggesting non-antipodal caps and a complex non-dipole magnetic field structure.

The orbital characteristics of \fourU\ are rather well determined, with a clearly established orbital period $P_{\rm orb}$ of 3.73~d as outlined by  \citet{2000ApJ...542L.131C}. 
Since then, the orbital period has been very precisely measured with
a marginally significant decay at a rate of ${\dot{P}}_{\rm orb}/ P_{\rm orb}\approx -10^{-6}$ yr$^{-1}$ \citep{2019ApJ...873...62H}.
Noteworthy, the work of \citet{1995A&A...303..497V} provided estimates on the orbital parameters and the mass function of the system, and unveiled a surprisingly low NS mass of about 1.0$\msun$.
Being one of the eclipsing binaries, \fourU\ also has a well-constrained orbital inclination  $i_{\rm orb}=67\degr \pm 1\degr$ \citep{2015A&A...577A.130F}.

The remainder of this paper is structured as follows. 
We present the observations and data reduction methods used in the study in Sect.~\ref{sec:Obs}. 
The results, including the pulse profile, spectral and polarimetric analysis are shown in Sect.~\ref{sec:results}. 
We discuss the results focusing on the polarization spectrum and pulsar geometry in Sect.~\ref{sec:discussion}. 
Finally, Sect.~\ref{sec:summary} summarizes the findings and the implications.

\section{Observations and data reduction}
\label{sec:Obs}

\begin{figure}%[h]
\centering
\includegraphics[width=\columnwidth]{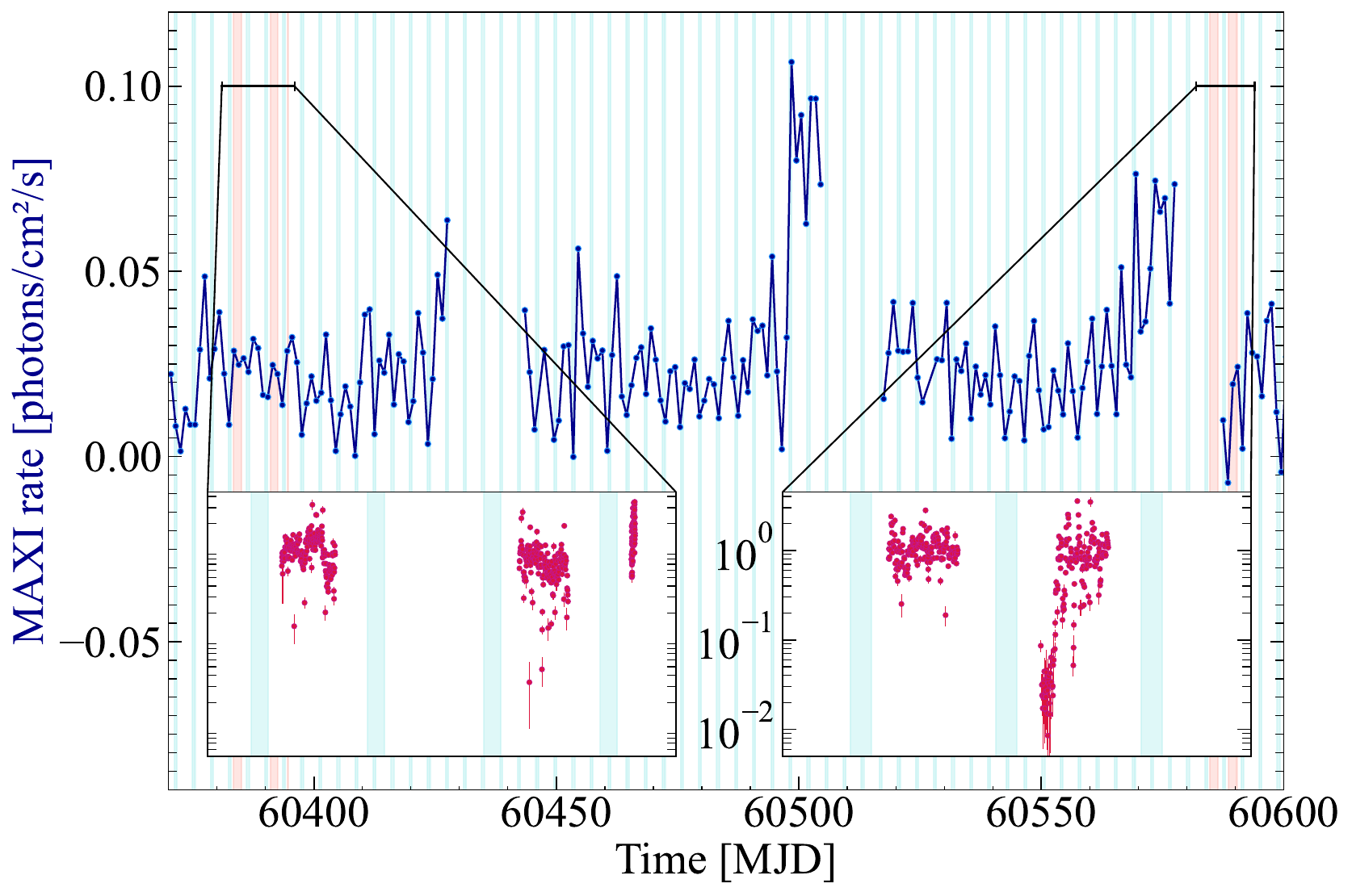}
\caption{One-day averaged light curve obtained by the MAXI monitor (4--10 keV) during the \ixpe\ observations of the source (dark blue points). 
The blue-shaded areas correspond to eclipse periods according to \citet{2015A&A...577A.130F}. 
The red shaded areas represent the \ixpe\ observations with the corresponding count-rate light curves (2--8 keV) shown in zoomed-in panels.}
\label{fig:MAXI}
\end{figure}

The \ixpe\ observatory, a NASA mission in partnership with the Italian Space Agency (ASI) \citep[see a detailed description in][]{Weisskopf2022}, has provided X-ray polarimetric data  on astrophysical X-ray sources since its launch in December 2021. 
It is equipped with three polarization-sensitive detector units \citep[DUs;][]{2021AJ....162..208S,2021APh...13302628B,DiMarco22b}, which provide polarimetry in the 2--8~keV energy band and timing with about a 10\,$\mu$s resolution. 
\fourU was observed by \ixpe\ five separate times during 2024 with a total of 353~ks of effective exposure. The first three observations occurred between March 14 and 25 and the remaining two between October 1 and 5 (see the observation log in Table~\ref{table:obs_log}).
During the second and third observations, a malfunction in the \ixpe\ data transmission system caused partial corruption of the data.
Due to this, only 7~ks of the third observation could be processed.  
Shortly after the third observation, the source exited the visibility window of \ixpe\ for several months. 
As a result, the dataset is widely spaced in time and does not cover consecutive orbits.
In addition, we utilize data from the Monitor of All-sky X-ray Image  \citep[{MAXI};][]{Matsuoka2009}. 
We show the \ixpe\ and MAXI light curves with times of orbital eclipses in Fig.~\ref{fig:MAXI}.

\begin{table}%[h]
\centering
\caption{Observations of \fourU\ by \ixpe.}
\label{table:obs_log}  
\begin{tabular}{cccc}
\hline\hline
Observation & ObsID & Epoch (MJD) & Exposure, ks \\%& Counts \\
\hline
1 & 03002701 & 60383.4--60385.1 & 90  \\%&  \\
2 & 03002801 & 60391.0--60392.5 & 77 \\%&  \\
3 & 03002901 & 60394.6--60394.7 & 7 \\%&  \\
4 & 03010401 & 60584.7--60586.5 & 92 \\%&  \\
5 & 03010301 & 60588.6--60590.3 & 87 \\%&  \\
\hline
\end{tabular}
\end{table}

The data have been processed with the \textsc{ixpeobssim} package version 31.0.3 using the CalDB released on 2024 February 28. 
Before the data analysis, the position offset correction and energy calibration were applied. 
Source photons were extracted using a circular region with a radius $R_{\rm src} = 80\arcsec$.
Due to the brightness of the source, we did not apply background subtraction and used the unweighted analysis \citep{DiMarco2022,DiMarco2023}.

Event arrival times were corrected to the barycenter of the solar system using the \texttt{barycorr} tool from the \textsc{ftools} package.
To account for the effect of binary motion, we corrected the observed event arrival times using the ephemerides from \citet{2015A&A...577A.130F}.
The spin period for each observation was searched for separately using $Z^2$ statistics.
The results were further refined using phase-connection, by fitting pulse arrival times.

Stokes $I$ spectra have been re-binned to have a minimum of 30 counts per energy channel, with the same binning also applied to the Stokes $Q$ and $U$ spectra.
The energy spectra were fitted using the \textsc{xspec} package (version 12.14.0) \citep{xspec1996} using $\chi^{2}$ statistics and version 13 instrument response functions (ixpe:obssim20240101:v13 for observations 1--3, and ixpe:obssim20240701:v13 for observations 4--5).
The reported uncertainties are given at the 68.3\% (1$\sigma$) confidence level (c.l.)  unless stated otherwise.

\section{Results}  
\label{sec:results} 

\subsection{Light curve}

The X-ray light curve of \fourU\ obtained by  MAXI\footnote{\url{http://maxi.riken.jp/}}  is displayed in Fig.~\ref{fig:MAXI}, with the 2--8 keV light curves obtained with \ixpe\ displayed in the insets.
The five \ixpe\ observations are largely consistent with each-other in terms of the count rate.
During the fifth observation (ObsID 03010301), a dip was observed following the eclipse. 
The data from the dip were excluded from all subsequent analysis.
The values of the spin period for the five observations are $P_{\rm spin}=526.26\pm0.09$~s, $526.29\pm0.09$~s, $525.7\pm1.7$~s, $526.21\pm0.08$~s, and $526.22\pm0.10$~s.

\subsection{Polarimetric analysis}

%%%%%%%%%%%%%%%%%%%%%%%%%
\begin{table}
\centering
\caption{Normalized Stokes parameters $q$ and $u$, PD, and PA for the phase-averaged data of \fourU for Obs.1--5 using the \texttt{pcube} algorithm. 
}
\begin{tabular}{lcccc}
    \hline\hline
    Interval  & $q$ & $u$ & PD & PA \\ 
          & (\%) & (\%) & (\%) & (deg) \\
    \hline
    Obs. 1 & $3.1\pm2.1$ & $\phantom{-}0.2\pm2.1$ & $3.1\pm2.1$ & $2\pm19$ \\
    Obs. 2--3 & $0.3\pm2.4$ & $-3.1\pm2.4$ & $3.1\pm2.4$ & $-43\pm22$ \\
    Obs. 4 & $2.4\pm2.0$ & $-1.3\pm2.0$ & $2.7\pm2.0$ & $-15\pm21$ \\
    Obs. 5\tablefootmark{a} & $4.0\pm2.5$ & $-3.6\pm2.5$ & $5.4\pm2.5$ & $-21\pm13$ \\
            \hline
    Obs. 1--5\tablefootmark{a} & $2.5\pm1.1$ & $-1.7\pm1.1$ & $3.0\pm1.1$ & $-18\pm11$ \\
            \hline    
    \end{tabular}
\tablefoot{Uncertainties are given at the 68.3\% (1$\sigma$) c.l. 
\tablefoottext{a}{Excluding dip.}}
\label{table:avebins}
\end{table}
%%%%%%%%%%%%%%%%%%%%%%%%%%

The polarimetric analysis of \fourU was performed using the \texttt{pcube} algorithm from the \texttt{xpbin} tool, which is implemented in the \textsc{ixpeobssim} package according to the formalism by \citet{2015-Kislat}.
We computed the normalized Stokes $q=Q/I$ and $u=U/I$ parameters and the PD using the equation $\mathrm{PD}=\sqrt{q^2+u^2}$, and the PA using $\mathrm{PA}=\frac{1}{2}\arctan\!2(u,q)$, with the PA measured from north to east counterclockwise on the sky.

\begin{figure} 
\centering
\includegraphics[width=0.85\columnwidth]{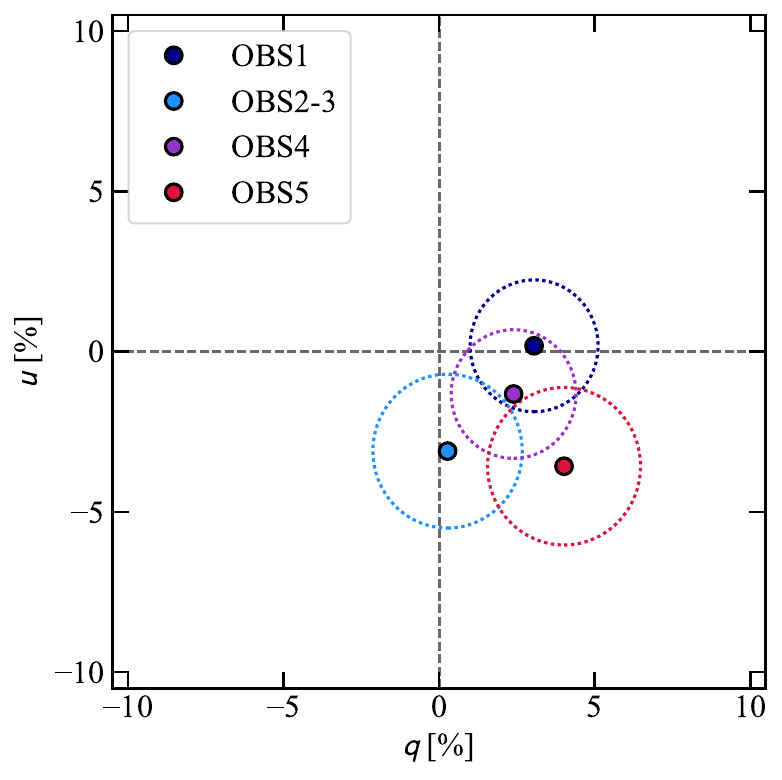}
\caption{Phase-averaged normalized Stokes $q$ and $u$ parameters for the individual observations in the full 2--8 keV energy range, obtained using the \texttt{pcube} algorithm. The size of the circles correspond to the uncertainty at a 68\% c.l.}
\label{fig:QU}
\end{figure}

The phase-averaged PD and PA in the entire 2--8 keV energy range of \ixpe\ for Obs. 1--5 are given in Table~\ref{table:avebins} and displayed in the normalized Stokes $q$ and $u$ plane in Fig.~\ref{fig:QU}.
Obs. 2 and 3 are combined into a single data set (Obs. 2--3).
During Obs. 5 we observed a dip, which we excluded from all subsequent data analysis.
We have not detected any significant polarization in the individual observations.

Because of the overall similarity between the individual observations of \fourU, and to achieve increased statistics, we combined the data from all five observations according to the following steps.
First, we determined the spin period for each individual observation and created pulse profiles.
These pulse profiles were then cross-correlated in order to determine the phase-shifts between the observations, which allowed us to connect phases correctly.
Finally, the events of the individual observations were phase-tagged and the data from all five observations combined into a single data set (excluding the dip during Obs. 5).
The phase-averaged PD and PA for the combined data set measured in the full 2--8~keV energy band are $3.0\pm1.1$\% and $-18\degr\pm11\degr$, respectively.

\begin{figure}%[h]
\centering
\includegraphics[width=\columnwidth]{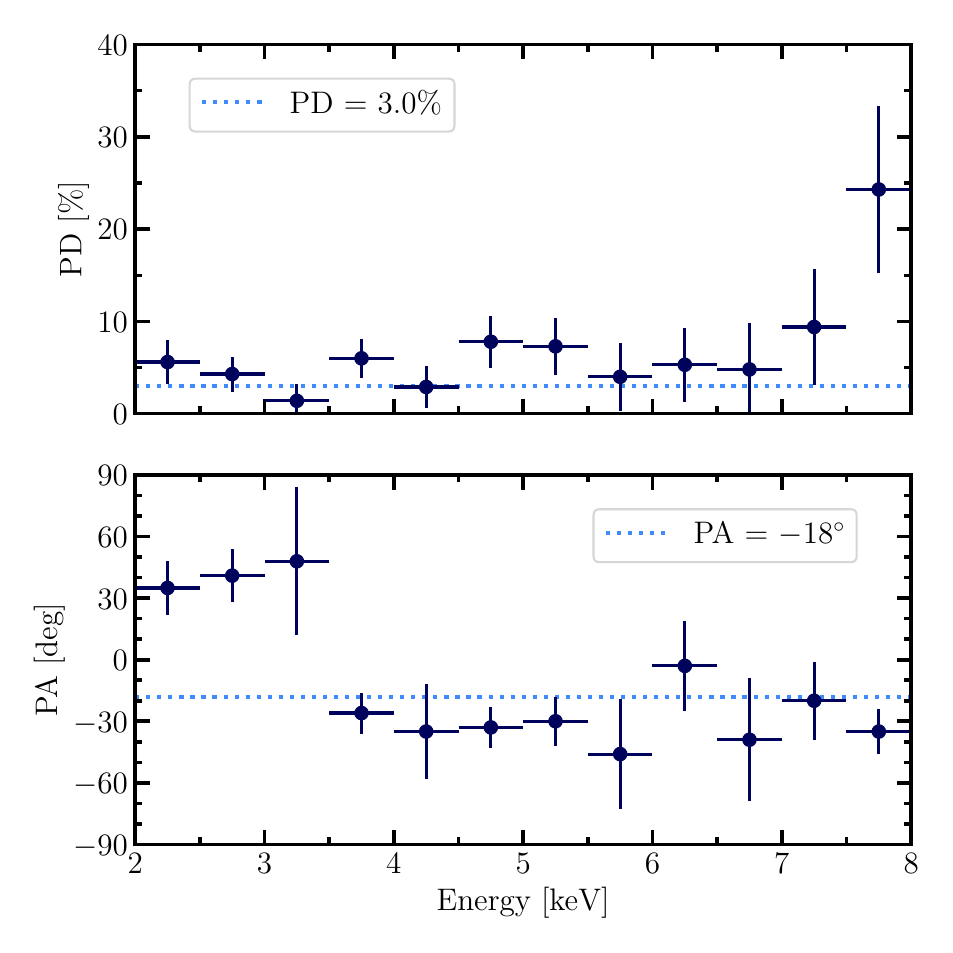}
\caption{Energy dependence of the PD and PA for the combined set of data (excluding dip), obtained with the \texttt{pcube} algorithm.}
\label{fig:enres-pcube}
\end{figure}

We also studied the energy dependence of the polarization properties by performing an energy-resolved polarimetric analysis dividing the combined data set into 12 energy bins.
The results of the energy-resolved analysis are given in Table~\ref{table:ebins} and displayed in Fig.~\ref{fig:enres-pcube}.
We see a clear energy dependence of the PA, with an about 70\degr\ span of the PA values between low and high energies.
A similar energy dependence of the polarization properties was first discovered in Vela~X-1 \citep{Forsblom23}, with a 90\degr\ switch in PA between low and high energies. 
The energy-dependent nature of Vela~X-1's polarization properties was further examined by \citet{Forsblom25}.

\begin{figure*} 
\centering
\includegraphics[width=0.9\columnwidth]{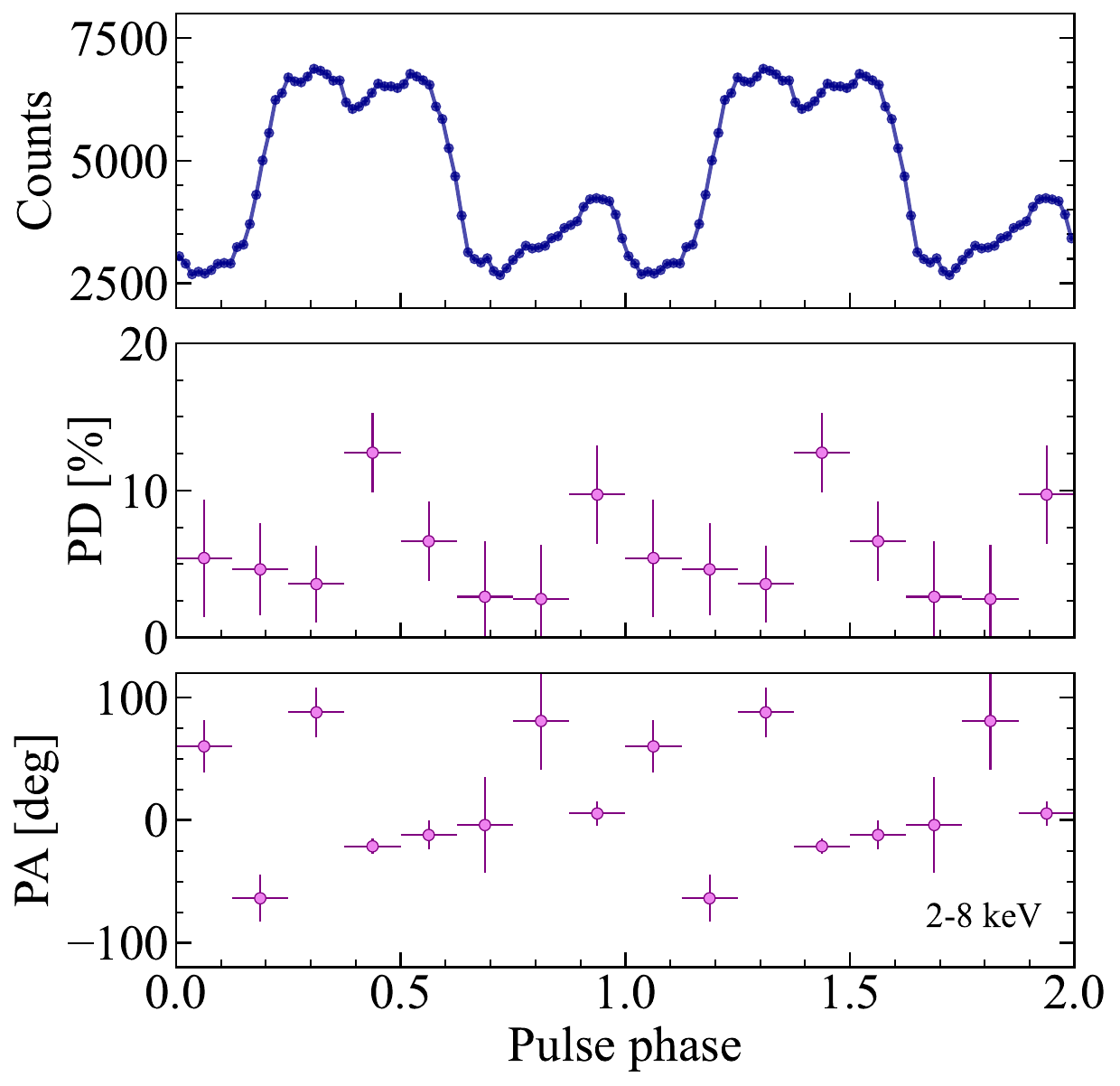}
\includegraphics[width=0.9\columnwidth]{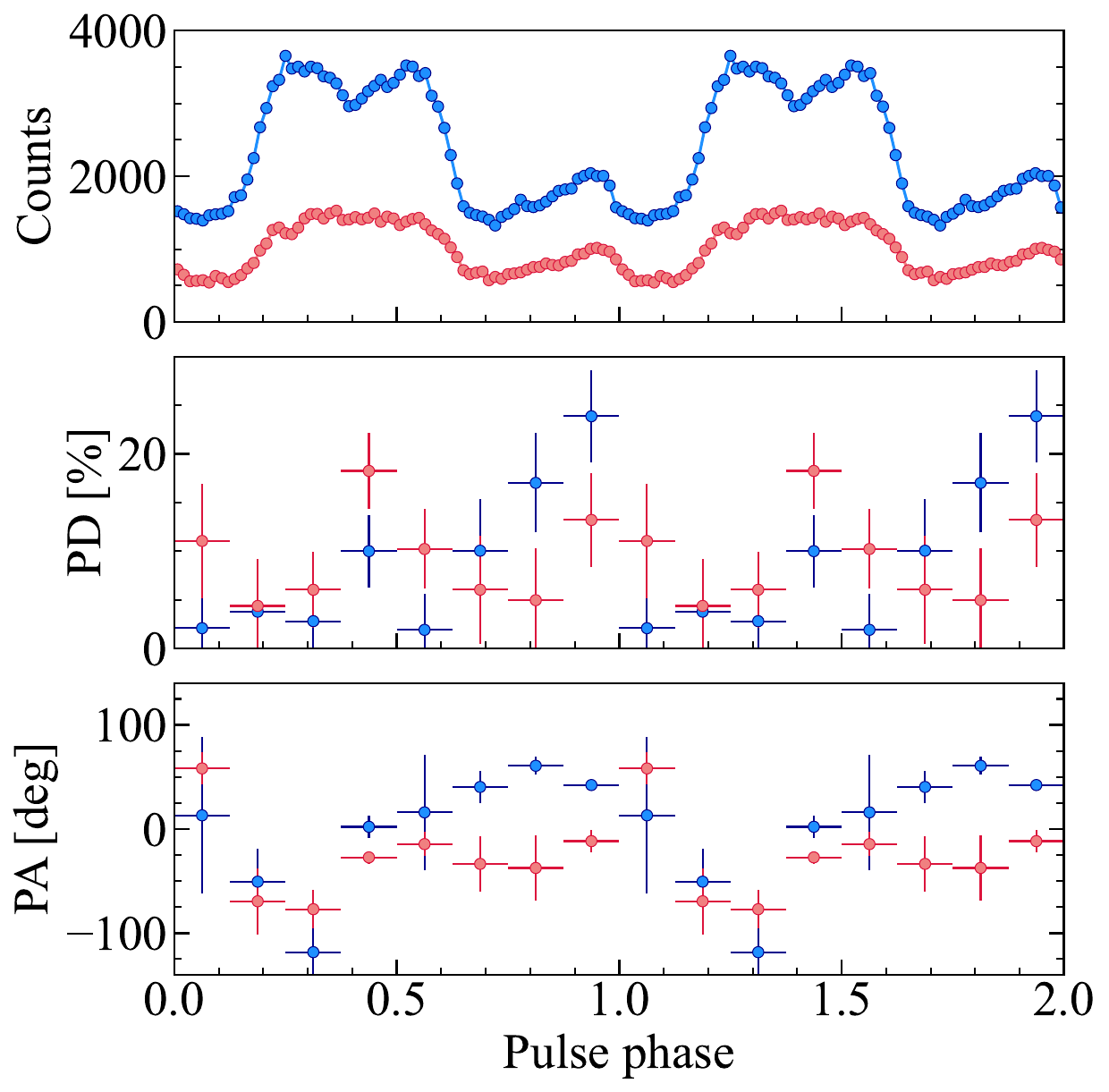}
\caption{Results of the phase-resolved polarimetric analysis of 4U 1538$-$52 using the \texttt{pcube} algorithm. Pulse profiles in units of counts per phase bin are displayed in the top panels, while the PD and PA in the central and lower panels. 
The left panels shows the results in the full \ixpe\ energy range 2--8 keV, while the right panels show the results separately for the low (2--3 keV) and the high (4--8 keV) energy ranges in blue and red, respectively.
The error bars correspond to one-sigma uncertainty. }
\label{fig:phase-res}
\end{figure*}

%%%%%%%%%%%%%%%%%%%%%%%%%
\begin{table}
\centering
\caption{Measurements of the normalized Stokes parameters $q$ and $u$, PD, and PA in different energy bins for the combined data set using the \texttt{pcube} algorithm. 
}
\begin{tabular}{ccccc}
    \hline\hline
    Energy  & $q$ & $u$ & PD & PA \\ %& $\mathrm{MDP_{99}}$ \\
       (keV)   & (\%) & (\%) & (\%) & (deg) \\ % & (\%) \\
    \hline
    2.0--2.5 & $1.9\pm2.4$ & $\phantom{-}5.2\pm2.4$ & $5.6\pm2.4$ & $35\pm 13$ \\
    2.5--3.0 & $0.5\pm1.9$ & $\phantom{-}4.2\pm1.9$ & $4.3\pm1.9$ & $41\pm 13$ \\
    3.0--3.5 & $-0.1\pm1.8$ & $\phantom{-}1.4\pm1.8$ & $1.4\pm1.8$ & $48\pm 36$ \\
    3.5--4.0 & $3.7\pm2.1$ & $-4.8\pm2.1$ & $6.0\pm2.1$ & $-26\pm 10$ \\
    4.0--4.5 & $1.0\pm2.3$ & $-2.7\pm2.3$ & $2.9\pm2.3$ & $-35\pm 23$ \\
    4.5--5.0 & $3.2\pm2.8$ & $-7.1\pm2.8$ & $7.8\pm2.8$ & $-33\pm 10$ \\
    5.0--5.5 & $3.7\pm3.1$ & $-6.2\pm3.1$ & $7.3\pm3.1$ & $-30\pm 12$ \\
    5.5--6.0 & $-0.1\pm3.7$ & $-4.0\pm3.7$ & $4.0\pm3.7$ & $-46\pm 27$ \\
    6.0--6.5 & $5.2\pm4.0$ & $-0.6\pm4.0$ & $5.3\pm4.0$ & $-3\pm 22$ \\
    6.5--7.0 & $0.9\pm5.0$ & $-4.7\pm5.0$ & $4.8\pm5.0$ & $-39\pm 30$ \\
    7.0--7.5 & $7.3\pm6.3$ & $\phantom{-}6.0\pm6.3$ & $9.4\pm6.3$ & $-20\pm 19$ \\
    7.5--8.0 & $8.3\pm9.0$ & $-22.9\pm9.0$ & $24.3\pm9.0$ & $-35\pm 11$ \\
    %2--3 & $1.3\pm1.5$ & $\phantom{-}4.8\pm1.5$ & $4.9\pm1.5$ & $37.5\pm 8.8$ \\
    %3--4 & $1.6\pm1.4$ & $-1.4\pm1.4$ & $2.1\pm 1.4$ & $-20.6\pm 18.6$ \\
    %4--5 & $2.0\pm1.8$ & $-4.7\pm1.8$ & $5.1\pm 1.8$ & $-33.6\pm 10.2$ \\
    %5--6 & $2.0\pm2.4$ & $-5.2\pm2.4$ & $5.6\pm 2.4$ & $-34.5\pm 12.1$ \\
    %6--7 & $3.2\pm3.2$ & $-2.6\pm3.2$ & $4.1\pm 3.2$ & $-19.3\pm 22.2$ \\
    %7--8 & $7.8\pm5.5$ & $-8.3\pm5.5$ & $11.4\pm 5.5$ & $-23.4\pm 13.8$ \\
    \hline
     2--8 & $2.5\pm1.1$ & $-1.7\pm1.1$ & $3.0\pm1.1$ & $-18\pm11$ \\
    \hline
    \end{tabular}
    \tablefoot{The uncertainties are given at the 68.3\% (1$\sigma$) c.l. }
    \label{table:ebins}
\end{table}

%%%%%%%%%%%%%%%%%%%%%%%%%%

Considering the importance of the PD and PA pulse-phase dependence, we performed a phase-resolved analysis by dividing the data of the combined set into eight uniform phase bins and used the \texttt{pcube} algorithm to determine polarimetric properties in each phase bin.
The results in the full 2--8~keV energy range are shown in Fig.~\ref{fig:phase-res} (left).
Considering the results of the energy-resolved analysis of the combined data set, we follow the analysis methods of \citet{Forsblom25} and analyze the phase-resolved polarization properties separately for the low (2--3 keV) and high (4--8 keV) energy bands, excluding the 3--4 keV energy range from the analysis in order to minimize the contribution from the adjacent energy band with different PA.
The results are displayed in Fig.~\ref{fig:phase-res} (right).
Normalized Stokes $q$ and $u$ parameters in the 2--8, 2--3, and 4--8 keV energy bands can be found in Table~\ref{table:best-fit-phaseres}.
 
%%%%%%%%%%%%%%%%%%%%%%%%%%
%%%%%%%%%%%%%%%%%%%%%%%%%
\begin{table*} 
\centering
\caption{Spectro-polarimetric parameters in different pulse-phase bins for the combined data set Obs. 1--5 in different energy ranges. }
\begin{tabular}{cccccccc}
    \hline\hline
    Phase & $q$ & $u$ & $N_{\mathrm{H}}$  & Photon index & PD & PA & $\chi^2$/d.o.f. \\ % 
           & (\%) & (\%) & ($10^{22}\mathrm{\;cm^{-2}}$) &   & (\%) & (deg) &  \\ % 
    \hline
    \multicolumn{8}{c}{\multirow{2}{*}{2--8 keV}} \\
    \\
    \hline
    0.000--0.125 & $-2.7\pm4.0$ & $4.7\pm4.0$ & $1.7\pm0.3$ & $1.09\pm0.07$ & $3.1\pm2.8$ & $-88\pm26$ & 1850/1742 \\
    0.125--0.250 & $-2.8\pm3.1$ & $-3.7\pm3.1$ & $2.5\pm0.2$ & $1.43\pm0.06$ & $2.5\pm2.3$ & $-69\pm32$ & 2574/2381 \\   
    0.250--0.375 & $-3.6\pm2.6$ & $-0.3\pm2.6$ & $2.1\pm0.2$ & $1.20\pm0.04$ & $2.2\pm1.9$ & $88\pm30$ & 2925/2794 \\  
    0.375--0.500 & $9.2\pm2.7$ & $-8.5\pm2.7$ & $1.6\pm0.2$ & $0.94\pm0.04$ & $8.1\pm1.9$ & $-16\pm7$ & 2814/2868 \\  
    0.500--0.625 & $6.0\pm2.7$ & $-2.7\pm2.7$ & $2.3\pm0.2$ & $1.27\pm0.05$ & $3.5\pm1.9$ & $1\pm17$ & 2956/2760 \\   
    0.625--0.750 & $2.8\pm3.8$ & $-0.4\pm3.8$ & $1.8\pm0.3$ & $1.00\pm0.07$ & $4.3\pm2.8$ & $15\pm20$ & 1973/1904 \\  
    0.750--0.875 & $-2.5\pm3.7$ & $0.8\pm3.7$ & $1.9\pm0.3$ & $1.00\pm0.07$ & $4.0\pm2.6$ & $86\pm20$ & 1999/1983 \\
    0.875--1.000 & $9.5\pm3.4$ & $1.8\pm3.4$ & $2.6\pm0.3$ & $1.04\pm0.06$ & $6.5\pm2.4$ & $7\pm11$ & 2192/2234 \\ 
    \hline
    \multicolumn{8}{c}{\multirow{2}{*}{2--3 keV}} \\
    \\
    \hline
%    Phase &  $N_{\mathrm{H}}$  & Photon index & PD & PA & $\chi^2$/d.o.f. \\ %& $q$ & $u$ 
%           & ($10^{22}\mathrm{\;cm^{-2}}$) &   & (\%) & (deg) &  \\ %& (\%) & (\%) 
%    \hline
    0.000--0.125 & $1.9\pm5.4$ & $0.9\pm5.4$ & $1.7^{\mathrm{fixed}}$ & $0.97\pm0.10$ & $2.3_{-2.3}^{+5.5}$ & $21\pm90$ & 726/700 \\
    0.125--0.250 & $-0.7\pm4.1$ & $-3.7\pm4.1$ & $2.5^{\mathrm{fixed}}$ & $1.35\pm0.08$ & $2.8_{-2.8}^{+4.1}$ & $-16\pm90$ & 908/878 \\   
    0.250--0.375 & $-1.5\pm3.6$ & $2.3\pm3.6$ & $2.1^{\mathrm{fixed}}$ & $1.16\pm0.07$ & $3.2_{-3.2}^{+3.5}$ & $51\pm90$ & 857/892 \\  
    0.375--0.500 & $10.0\pm3.7$ & $0.7\pm3.7$ & $1.6^{\mathrm{fixed}}$ & $0.96\pm0.07$ & $7.8\pm3.7$ & $8\pm14$ & 827/906 \\  
    0.500--0.625 & $1.6\pm3.7$ & $1.0\pm3.7$ & $2.3^{\mathrm{fixed}}$ & $1.24\pm0.07$ & $1.6_{-1.6}^{+3.7}$ & $70\pm90$ & 898/899 \\   
    0.625--0.750 & $1.6\pm5.3$ & $9.9\pm5.3$ & $1.8^{\mathrm{fixed}}$ & $0.92\pm0.10$ & $9.1\pm5.3$ & $26\pm18$ & 765/769 \\  
    0.750--0.875 & $-8.9\pm5.1$ & $14.5\pm5.1$ & $1.9^{\mathrm{fixed}}$ & $0.92\pm0.10$ & $17.6\pm5.1$ & $65\pm9$ & 779/776 \\
    0.875--1.000 & $2.3\pm4.7$ & $23.8\pm4.7$ & $2.6^{\mathrm{fixed}}$ & $1.01\pm0.09$ & $17.6\pm4.7$ & $45\pm8$ & 774/826 \\ 
    \hline
    \multicolumn{8}{c}{\multirow{2}{*}{4--8 keV}} \\
    \\
    \hline
%    Phase &  $N_{\mathrm{H}}$  & Photon index & PD & PA & $\chi^2$/d.o.f. \\ %& $q$ & $u$ 
%           & ($10^{22}\mathrm{\;cm^{-2}}$) &   & (\%) & (deg) &  \\ %& (\%) & (\%) 
%    \hline
    0.000--0.125 & $-4.9\pm5.9$ & $9.9\pm5.9$ & $1.7^{\mathrm{fixed}}$ & $0.87\pm0.10$ & $5.2\pm4.8$ & $-88\pm27$ & 503/460 \\
    0.125--0.250 & $-3.3\pm4.8$ & $-2.9\pm4.8$ & $2.5^{\mathrm{fixed}}$ & $1.19\pm0.08$ & $3.4_{-3.4}^{+3.8}$ & $83\pm90$ & 674/667 \\   
    0.250--0.375 & $-5.4\pm3.9$ & $-2.6\pm3.9$ & $2.1^{\mathrm{fixed}}$ & $1.02\pm0.06$ & $8.0\pm3.0$ & $-53\pm11$ & 1023/946 \\  
    0.375--0.500 & $10.6\pm3.9$ & $-14.9\pm3.9$ & $1.6^{\mathrm{fixed}}$ & $0.77\pm0.06$ & $11.5\pm3.0$ & $-29\pm8$ & 1031/1036 \\  
    0.500--0.625 & $8.9\pm4.1$ & $-5.0\pm4.1$ & $2.3^{\mathrm{fixed}}$ & $1.09\pm0.07$ & $7.0\pm3.2$ & $4\pm14$ & 1032/955 \\   
    0.625--0.750 & $2.4\pm5.6$ & $-5.6\pm5.6$ & $1.8^{\mathrm{fixed}}$ & $0.91\pm0.10$ & $3.9_{-3.9}^{+4.6}$ & $9\pm90$ & 549/514 \\  
    0.750--0.875 & $1.3\pm5.4$ & $-4.8\pm5.4$ & $1.9^{\mathrm{fixed}}$ & $0.70\pm0.09$ & $5.1\pm4.3$ & $-42\pm29$ & 542/541 \\
    0.875--1.000 & $12.1\pm4.8$ & $-5.2\pm4.8$ & $2.6^{\mathrm{fixed}}$ & $1.03\pm0.08$ & $9.0\pm3.8$ & $-12\pm13$ & 643/667 \\ 
    \hline
    \end{tabular}
\tablefoot{ 
Normalized Stokes parameters $q$ and $u$ are obtained using the \texttt{pcube} algorithm in {\sc ixpeobssim}.
PD and PA are obtained with \textsc{xspec}.
The uncertainties  computed using the \texttt{error} command 
are given at the 68.3\% (1$\sigma$) c.l. ($\Delta\chi^2=1$ for one parameter of interest). } 
\label{table:best-fit-phaseres}
\end{table*}
%%%%%%%%%%%%%%%%%%%%%%%%%%

Next, we performed a full phase-averaged spectro-polarimetric analysis of the combined data set, in order to account for the spectral shape and the energy dispersion.
Source Stokes $I$, $Q$, and $U$ spectra were extracted using the \texttt{PHA1}, \texttt{PHA1Q}, and \texttt{PHA1U} algorithms from the \texttt{xpbin} tool, which produces full data sets of nine spectra, three per DU.
Stokes $I$, $Q$, and $U$ spectra were produced for each separate observation, and were fitted simultaneously with \textsc{xspec} (total number of spectra equal to 45).

%%%%%%%%%%%%%%%%%%%%%%%%%%
\begin{figure}
\centering
\includegraphics[width=0.8\linewidth]{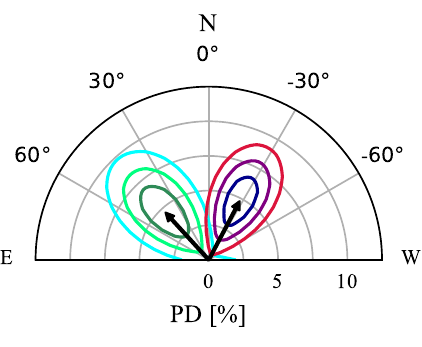}
\caption{Polarization vectors of \fourU from the results of the phase-averaged spectro-polarimetric analysis of the combined data set. Contours at 68.3\%, 95.45\%, and 99.73\% c.l. calculated for two degrees of freedom are shown in shades of blue for 2--3 keV and in shades of red for 4--8 keV.}
 \label{fig:phase-ave-xspec}
\end{figure}

The spectral continuum for \fourU can be described using a simple absorbed power-law model plus an iron line at $\sim$6.4~keV.
Considering the restrictions imposed by the \ixpe\ energy range and the energy resolution of the instrument \citep{Weisskopf2022}, we did not include the iron line in the final model used for the spectral fitting.
We adopted a \texttt{powerlaw} model, with interstellar absorption introduced using the \texttt{tbabs} model with abundances from \cite{Wilms2000}. The \texttt{polconst} polarization model, which assumes energy-independent PD and PA, was applied to the spectral continuum. The re-normalization constant, \texttt{const}, was introduced to account for possible discrepancies between the individual DUs, and was fixed to unity for DU1.
The final spectral model,
\begin{eqnarray*} 
\texttt{ tbabs$\times$polconst$\times$powerlaw$\times$const},
%\label{eq:fit}
\end{eqnarray*}
was applied to both the phase-averaged and phase-resolved data.
We allow for the power-law normalization to vary between the individual observations.
The initial spectral analysis was performed over the full \ixpe\ energy range of 2--8 keV.
Considering the results of the energy-resolved analysis using \texttt{pcube}, we also confined the fit to the low (2--3 keV) and high (4--8 keV) energy bands.
The \texttt{steppar} command in \textsc{xspec} was used to create confidence contours for both the phase-averaged and phase-resolved analysis.
Contour plots at 68.3\%, 95.45\%, and 99.73\% c.l. are shown in Fig.~\ref{fig:phase-ave-xspec} for the phase-averaged results at low (2--3 keV) and high (4--8 keV) energies.
The average PA difference between these components is about 70\degr.
Spectral parameters for the best-fit models from the results of the phase-averaged spectro-polarimetric analysis of the combined data set are listed in Table~\ref{table:best-fit}.

%%%%%%%%%%%%%%%%%%%%%%%%%%
\begin{figure*}
\centering
\includegraphics[width=0.24\linewidth]{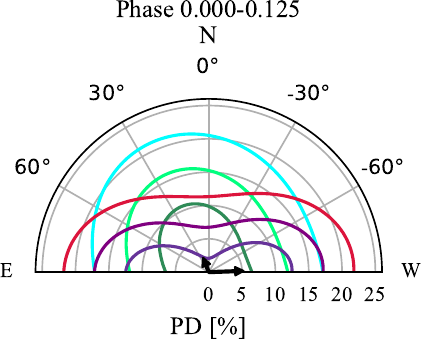}
\includegraphics[width=0.24\linewidth]{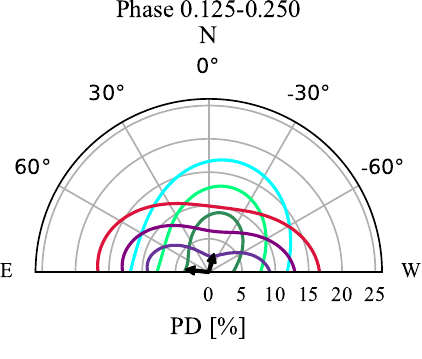}
\includegraphics[width=0.24\linewidth]{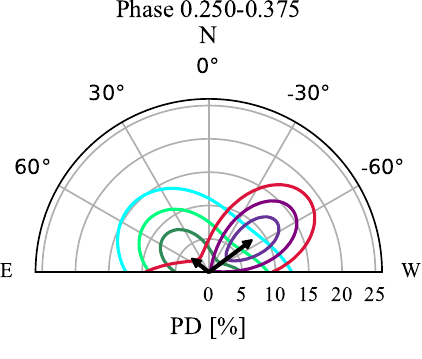}
\includegraphics[width=0.24\linewidth]{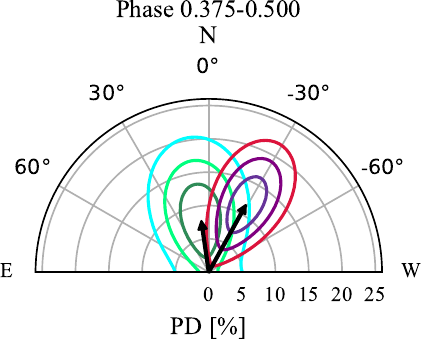}
\includegraphics[width=0.24\linewidth]{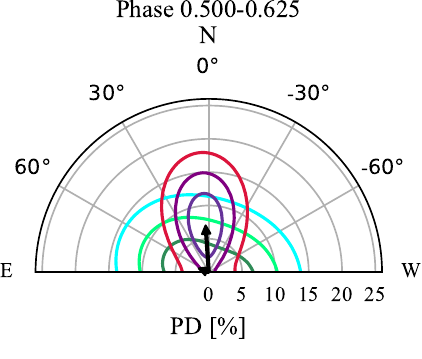}
\includegraphics[width=0.24\linewidth]{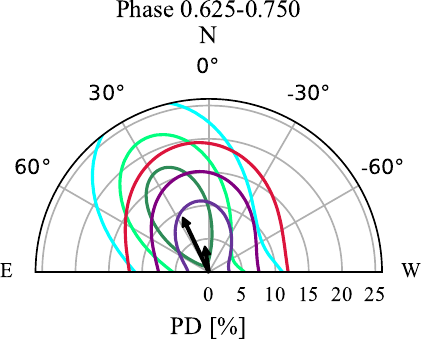}
\includegraphics[width=0.24\linewidth]{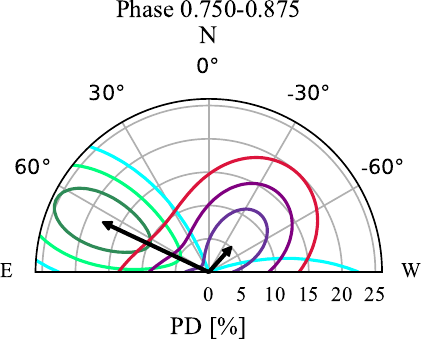}
\includegraphics[width=0.24\linewidth]{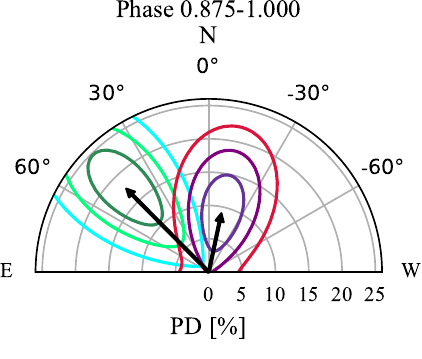}
\caption{Polarization vectors of \fourU from the results of the phase-resolved spectro-polarimetric analysis of the combined data set. Contours at 68.3\%, 95.45\%, and 99.73\% c.l. calculated for two d.o.f are shown in shades of blue  for 2--3 keV and shades of red for 4--8 keV.}
\label{fig:phase-res-xspec}
\end{figure*}

A phase-resolved spectro-polarimetric analysis was performed in a similar fashion. The data were separated into eight uniform phase-bins by extracting Stokes $I$, $Q$, and $U$ spectra for each individual phase bin using the \texttt{xpbin} tool's \texttt{PHA1}, \texttt{PHA1Q}, and \texttt{PHA1U} algorithms.
The same model was used for the phase-resolved spectral analysis as for the phase-averaged, with the cross-calibration constants for DU2 and DU3 set to the values of the phase-averaged analysis (see Table~\ref{table:best-fit}).
Similarly to the phase-averaged analysis, we also confined the spectral fitting to the low (2--3 keV) and high (4--8 keV) energy ranges considering the results of the energy-resolved \texttt{pcube} analysis.
The results of the phase-resolved spectro-polarimetric analysis for the combined data set in the 2--8, 2--3, and 4--8 keV energy ranges are given in Table~\ref{table:best-fit-phaseres}, with the corresponding contour plots at 68.3\%, 95.45\%, and 99.73\% c.l. shown in Fig.~\ref{fig:phase-res-xspec} for the low (2--3 keV) and high (4--8 keV) energies.

%%%%%%%%%%%%%%%%%%%%%%%%%%
\begin{table*}
\centering
\caption{Parameters of the best-fit spectro-polarimetric \textsc{xspec} model to the phase-averaged combined data set in different energy bands.
} 
\begin{tabular}{llccccc}
\hline \hline
Component & Parameter & Unit & 2--8 keV & 2--3 keV & 4--8 keV\\
\hline
\texttt{tbabs} & $N_{\mathrm{H}}$ & $10^{22}\mathrm{\;cm^{-2}}$ & $2.11\pm0.08$ & $2.11^{\mathrm{fixed}}$ & $2.11^{\mathrm{fixed}}$ \\
\texttt{powerlaw} & Photon index &  & $1.14\pm0.02$ & $1.11\pm0.03$ & $1.00\pm0.02$ \\
\texttt{polconst} & PD & \%      & $1.7\pm0.8$ & $4.6\pm1.5$ & $4.8\pm1.2$ \\
 & PA & deg     & $-9\pm14$ & $42\pm10$ & $-28\pm8$ \\
\texttt{constant} & $\mathrm{const_{DU2}}$ &  & $1.017\pm0.004$ & $1.016\pm0.007$ & $1.015\pm0.009$ \\
 & $\mathrm{const_{DU3}}$ &  & $1.003\pm0.004$ & $1.004\pm0.007$ & $0.998\pm0.009$ \\
\hline
 & $\mathrm{Flux_{2-8\;keV}}$\tablefootmark{\textbf{a}} & $10^{-10}$\,erg\,s$^{-1}$\,cm$^{-2}$  & $2.13\pm0.03$ & -- & -- \\
 & $L_{2-8\rm\,keV}$\tablefootmark{\textbf{b}} & $10^{35}$\,\,erg\,s$^{-1}$  & 9.14 & -- & -- \\
 & $\chi^2$ (d.o.f.) &  & 5545 (5029) & 1096 (1025) & 2935 (2843) \\
\hline  
\end{tabular}
\tablefoot{
The uncertainties are given at the 68.3\% (1$\sigma$) c.l. and were obtained using the \texttt{error} command in \textsc{xspec} with $\Delta\chi^2=1$ for one parameter of interest. 
\tablefoottext{\textbf{a}}{Observed flux in the 2–8 keV range. }
\tablefoottext{b}{Luminosity is computed using the distance of $d=6$~kpc \citep{Gaia-phot,2023-HMXBcat}.}
}
\label{table:best-fit}
\end{table*}

\section {Discussion}
\label{sec:discussion} 

\ixpe\ observations of accreting XRPs have revealed a multitude of details that challenge the theoretical predictions for these sources made before the \ixpe\ era. 
\fourU\ exhibits a relatively low PD, consistent with the broader trend across the XRPs detected with \ixpe\ to date.
This persistent pattern underscores the necessity for developing new theories that incorporate additional factors influencing polarization properties. 
More significantly, the unprecedented PA behavior highlights this source as a unique case. 
So far, the \ixpe\ mission has enabled the geometries of pulsars to be determined for the first time in several systems using the rotating vector model (RVM). 
We also analyze the phase-resolved polarization properties of \fourU\ using the RVM, overcoming the challenges posed by the energy-dependent PA.

\subsection{Pulsar geometry model} 
\label{sec:geom}

The RVM, as introduced by \citet{1969ApL.....3..225R} and \citet{1988ApJ...324.1056M}, relates the observed PA of linear polarization to the projection of the pulsar magnetic dipole on the sky. 
Modeling of the observed pulse phase-dependence of the PA using the RVM has been by now successfully applied to most of the XRPs observed by \ixpe. 
Assuming that radiation is dominated by ordinary mode (O-mode) photons, the PA $\chi$ as a function of the rotation phase $\phi$ is given by \citep{2020A&A...641A.166P}:
\begin{equation} \label{eq:rvm}
   \tan (\chi-\chi_{\rm p}) = \frac{-\sin \theta\ \sin (\phi-\phi_0)}
   {\sin i_{\rm p} \cos \theta  - \cos i_{\rm p} \sin \theta  \cos (\phi\!-\!\phi_0) } ,
\end{equation} 
where $\phi_0$ is the phase when the northern magnetic pole passes in front of the observer, 
$\chi_{\rm p}$ is the position angle of the pulsar angular momentum (as the polarization plane of the O-mode is aligned with the magnetic axis at $\phi_0$),
$i_{\rm p}$ is the inclination of the pulsar spin axis to the line of sight, 
$\theta$ is the magnetic obliquity (the angle between the spin and magnetic dipole axes).

\begin{figure*}%[h]
\centering
\includegraphics[width=0.85\textwidth]{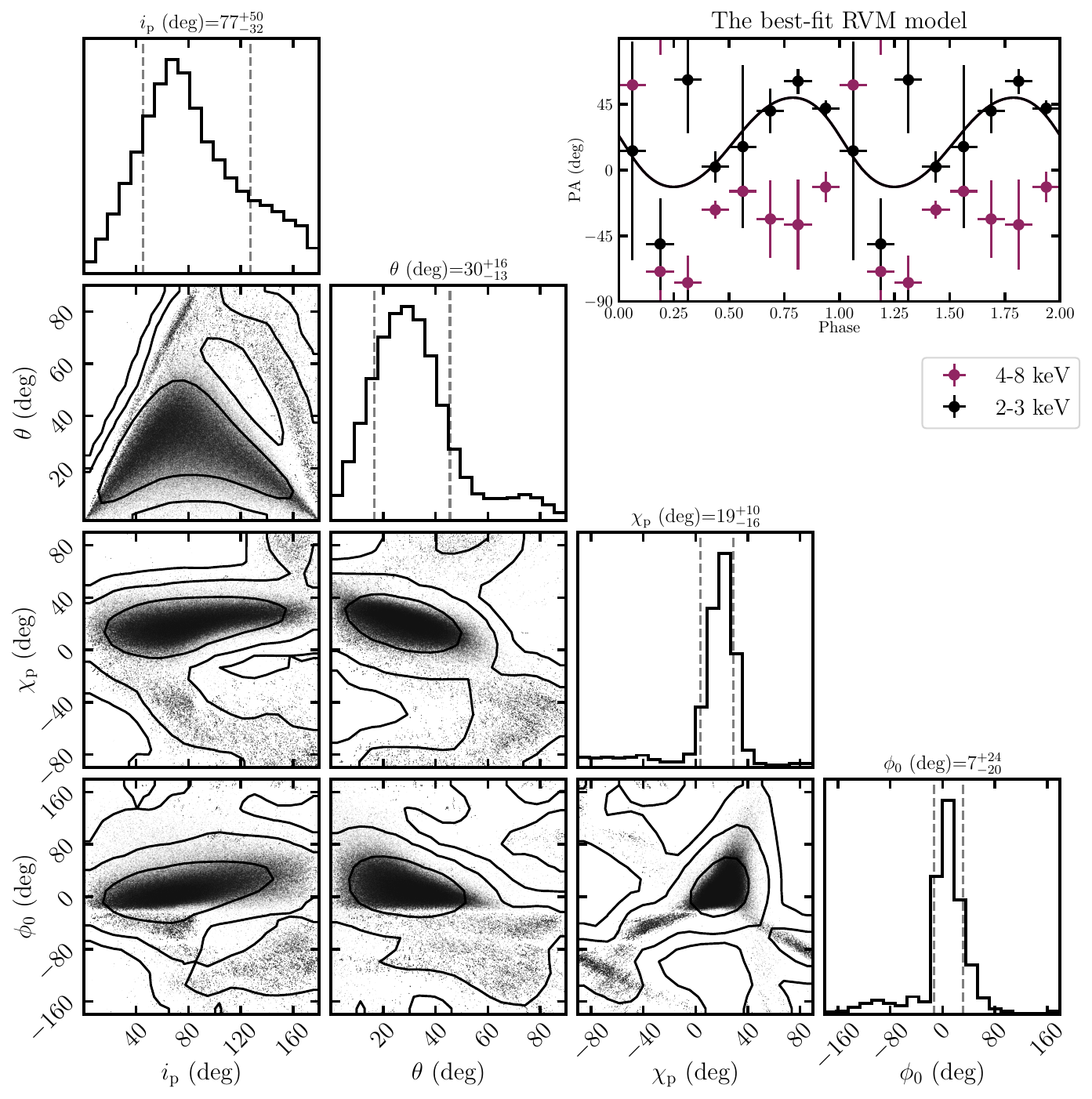}
\caption{Results of RVM fitting of the pulsar phase-resolved X-ray polarization properties in the 2--3 keV energy band presented in a corner plot. On the upper right panel, the PA data shown with circles and error bars denoting the bin widths and one-sigma errors are plotted against the best-fit RVM model (solid lines). The black color shows the base case fit to only the lower energy band and the magenta shows the higher energy data. }
\label{fig:RVMchi}
\end{figure*}
    
We follow the procedure outlined in \citet{2023A&A...678A.119S} and \citet{Forsblom24} to fit the RVM using  Markov chain Monte Carlo simulations. 
We use the affine-invariant ensemble sampler implemented in the {\sc Python} pachage {\sc emcee}  \citep{2013PASP..125..306F} to perform the fitting. 
Directly fitting these PA values can lead to biases, especially in low signal-to-noise regimes, as the PA is not normally distributed.
Instead, we work with the Stokes parameters in our fitting procedure using the appropriate likelihood function. 
For polarization data in one pulse phase bin, we use the probability density function of the PA as derived by \citet{1993A&A...274..968N}:
\begin{equation} \label{eq:PA_dist}
   G(\psi) = \frac{1}{\sqrt{\pi}} 
   \left\{  \frac{1}{\sqrt{\pi}}  + 
   \eta {\rm e}^{\eta^2} 
   \left[ 1 + {\rm erf}(\eta) \right]
   \right\} {\rm e}^{-p_0^2/2} , 
\end{equation}
where
$\eta=p_0 \cos[2(\psi-\psi_0)]/\sqrt{2}$,
$\psi$ is the model PA,  
$\psi_0=\frac{1}{2}\arctan(u/q)$ is the measured PA from Stokes parameters, and 
$p_0=\sqrt{q^2+u^2}/\sigma_{\rm p}$ is the measured PD relative to its error,
and \mbox{erf} is the  error function.
The log-likelihood for a set of phase bins is then just the sum of the log-likelihoods over all phase bins:
\begin{equation}
 \log L = \sum_i \log G(\psi_i).
\end{equation}
We note that this procedure allows us to use all phase bins, even those where polarization was not detected at a statistically significant level. 
The probability distribution of the PA is rather flat in these bins and does not affect the obtained RVM parameters.

First, we performed the MCMC fit of the RVM model on the data in the 2--3 keV energy band.  
We use the values for the normalized  Stokes parameters $q$ and $u$ in the 8 phase bins given in Table~\ref{table:best-fit-phaseres}.
The results of the fitting process and the best-fit model for the PA are shown in Fig.~\ref{fig:RVMchi}. 
The best-fit parameters are: $i_{\rm p} = 77^{+50}_{-32}$~deg, $\theta = 30^{+16}_{-13}$~deg, $\chi_{\rm p} = 19^{+10}_{-16}$~deg, and $\phi_0= 7^{+24}_{-20}$~deg. 
The inclination of the pulsar rotation axis $i_{\rm p}$ and the magnetic obliquity $\theta$ derived from the MCMC analysis are compatible with the corresponding values obtained from the cyclotron line region pulse-phase fits performed by \citet{Bulik1992}, in some of their models. 
The high $>$50\degr\ spin-axis inclination is also consistent with alignment to the binary orbital plane, which is known to be inclined at $67\degr \pm 1\degr$ \citep{2015A&A...577A.130F}.
Because the 2--3 keV energy band contains only a fraction of the total registered photons, the uncertainties on the constraints from RVM fitting are large.

The 4--8 keV energy band, in turn, contains about two times less photons than the 2--3 keV band. 
Fitting the RVM model to the high energy band data exclusively provides a rather unreliable fit with very weak constraints: 
$i_{\rm p} = 42^{+42}_{-27}$~deg, $\theta = 42^{+29}_{-30}$~deg, $\chi_{\rm p} = -34^{+87}_{-37}$~deg, and $\phi_0= 47^{+48}_{-56}$~deg.
The parameter spaces explored by the MCMC simulations for these two energy ranges show largely distinct probability clouds, with minimal overlap in their posterior distributions. 
We attempted to combine the datasets with a 90\degr\ turn in PA for the second dataset, but it became clear that this could not be justified. 
In fact, we attempted to fit both datasets simultaneously with a single RVM model, where all parameters were shared except for the angle $\chi_{\rm p}$. 
Albeit unphysical, we allowed for an offset $\Delta \chi_{\rm p}$ between the $\chi_{\rm p}$ values of the two datasets, treating this offset as a free parameter.
The best-fit value for $\Delta \chi_{\rm p}$ was found to be $44^{+16}_{-11}$ degrees, which is the furthest from the right angle.
Thus, a single geometric configuration struggles to simultaneously explain the polarization properties across both energy bands. 
In summary, it is clear that the observed polarization features of the pulse profile cannot be adequately explained by a simple RVM.
Possible explanations for the observed behavior are discussed in the next section.
%implying more complex magnetic field geometries or the presence of multiple polarized emission components.

\begin{figure}%[h]
\centering
\includegraphics[width=0.9\columnwidth]{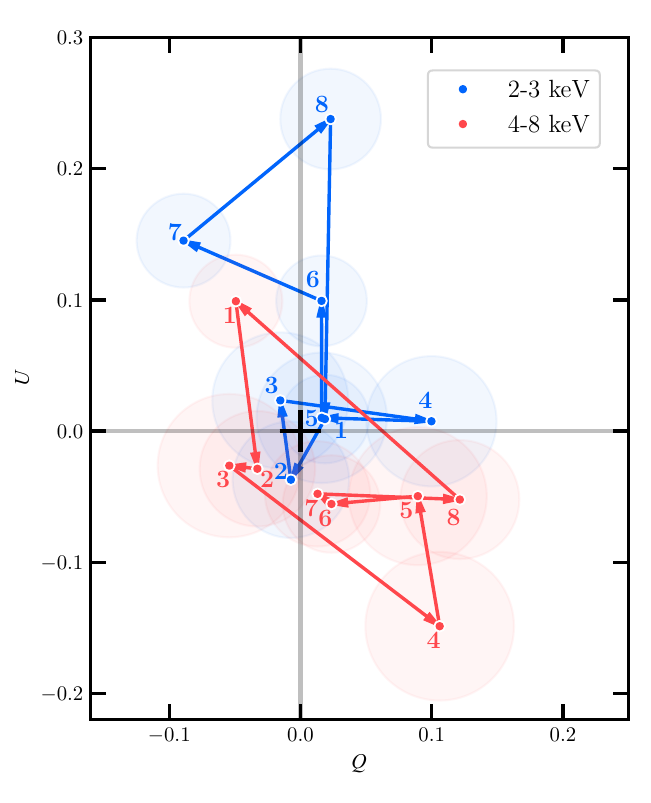}
\caption{Absolute Stokes parameters  in 8 phase bins shown for 2--3 keV band (in blue) and 4--8 keV band (in red) normalized by the average flux in each band. 
The circle sizes correspond to the uncertainties at 68\% ($1\sigma$) c.l.
The data points are annotated with numbers in accordance to Table~\ref{table:best-fit-phaseres}. }
\label{fig:absolute-qu}
\end{figure}

\subsection{Energy dependence of the PA}

As previously emphasized, the flip in the PA happens at roughly 3 keV.   
We note, that earlier study of \fourU\ with \textit{XMM-Newton} presented by \citet{2011A&A...526A..64R} shows evidence that a softer spectral component becomes dominant below this energy.  
Similar spectral dependence of the polarization parameters, including the PA flip between spectral components, has previously been observed for another XRP, namely \mbox{Vela~X-1} \citep{Forsblom23}. 
\mbox{Vela~X-1} displays a unique 90\degr\ switch in PA between low and high energies, with the change in PA appearing at 3.4~keV. The PD shows a gradual decrease at lower energies, reaching zero at 3.4 keV (corresponding to the 90\degr\ PA switch), then gradually increasing at higher energies.
The energy dependence of \mbox{Vela~X-1} was examined in more detail in \citet{Forsblom25}, where the pulse-phase dependence of the polarimetric properties similarly was analyzed separately for low and high energies. Two different scenarios were introduced in order to explain the 90\degr\ difference between low and high energies: a two-component spectral model and vacuum resonance. 
\fourU and \mbox{Vela~X-1} are very similar systems, with long spin periods, short orbital periods and stellar companions of spectral type B0--0.5 I. Both systems also display a so-called soft excess.
Considering the similarities not only between the systems themselves, but also the energy-dependent nature of their polarization properties as discovered by \ixpe, it is very likely that the same underlying phenomena is responsible for the energy-dependence. 

However, there is a stark difference between \fourU and \mbox{Vela~X-1} in terms of the PA switch between the energy bands. 
In \mbox{Vela~X-1}, the PA change strikingly agrees with a 90\degr\ shift in both the phase-averaged and phase-resolved analyses. 
In contrast, \fourU shows a phase-averaged PA switch of only about 70\degr, with the phase-resolved analysis revealing irregular variations, reaching approximately 40\degr\ in phase bin 4 (phases 0.375--0.5).
A 90\degr\ angle could have allowed for a simpler explanation \citep{Forsblom25}, but the irregular PA switch in \fourU\ is more challenging to interpret.

We also note that both pulsars, \fourU and \mbox{Vela~X-1}, have been studied for possible pole asymmetry in pulse profile analysis \citep{Bulik1995}. 
However, the local arrangement of magnetic poles on the pulsar surface could not satisfactorily explain the difference in the PA profile across the \ixpe\ energy range, since the assumption of the RVM is that the PA is determined at the adiabatic radius, where the field structure is dipolar \citep{Heyl2002,Taverna15,Heyl2018,Poutanen2024Galax}.
Consequently, an additional polarized component may be responsible for the difference in polarization behavior between the energy bands. 
A very similar approach has been implemented by \citet{Doroshenko23} for the Be-transient XRP \mbox{LS~V~+44~17} and by \citet{Poutanen2024} for the ultraluminous XRP \mbox{Swift~J0243+6124}, except that in those cases the additional polarized component varied between observations rather than with energy. 
We show the absolute Stokes parameters in two energy ranges in units of phase-averaged flux of each respective energy band depending on the phase of the pulsar on the $Q$--$U$ plane in Fig.~\ref{fig:absolute-qu}.
This shows that the evolution with phase in the two energy bands is drastically different, and the figures the pulsar traces on the $Q$--$U$ plane are markedly distinct. 
Thus, even if the difference can be explained by an additional polarized component, this component must depend on the pulsar phase.

We assume that near a NS, radiation propagates as normal modes with polarization either in the plane containing $B$-field and photon momentum or perpendicular to it. 
Within the adiabatic radius, the polarization of photons relative to the local magnetic field direction remains unchanged. 
Consequently, processes occurring within this region can only produce linear polarization at two PAs differing by 90\degr.
A polarization jump at an angle other than 90\degr\ suggests the influence of processes occurring beyond the adiabatic radius. 
A likely candidate is the scattering of X-ray photons by the magnetospheric accretion flow, which could introduce additional polarization with a PA that varies smoothly with pulse phase. However, at the observed luminosity level, the magnetospheric accretion flow is unlikely to scatter more than 1\% of the X-ray photons.
Other potential sources of polarized radiation beyond the adiabatic radius include scattering off the stellar wind of the companion \citep{Ahlberg24} or accretion disk wind \citep{Doroshenko23,Poutanen2024, Nitindala25}, or reflection from the  disk itself \citep{Matt93,Poutanen96}. 
These components of the X-ray flux could also contribute to phase-dependent polarization variations (see discussion in \citealt{Tsygankov22}).     

\section{Summary}
\label{sec:summary} 

\fourU was observed by \ixpe\ on five separate occasions, three times in March 2024, and twice in October 2024.
The results of the first polarimetric study of \fourU can be summarized as follows:
\begin{enumerate}
\item We do not detect significant polarization in any of the five  observations separately in the pulse phase-averaged data. However, we observe similar polarization properties across the individual observations. Combining the data from all observations gives a marginally significant detection of polarization, with PD=$3.0\pm1.1$\% and PA=$-18\degr\pm11\degr$.
\item The energy-resolved analysis reveals a substantial energy dependence of the polarization properties in \fourU. 
The pulse phase-averaged analysis shows a remarkable $\sim$70\degr\ switch in PA between low and high energies, occurring around 3 keV, with indications of drastically different PA behavior with the pulse phase for the lower and higher energy ranges. 
This kind of energy-dependent polarization is unprecedented in XRPs, having previously been observed only in \mbox{Vela~X-1}, where a 90\degr\ shift was detected in both phase-averaged and phase-resolved analyses. 
However, the $\sim$70\degr\ rotation in \fourU\ presents a more significant theoretical challenge.

\item The spectro-polarimetric analysis of the phase-averaged data confirms the energy-dependence, where for the low- and high energy bands we detect significant polarization, with PD=$4.6\pm1.5$\% and PA=$42\degr\pm10\degr$ in 2--3 keV, and PD=$4.8\pm1.2$\% and PA=$-28\degr\pm8\degr$ in 4--8 keV bands.

\item The pulse phase-resolved polarimetric analysis similarly reveals different behavior in the low and high energy ranges. 
At low energies (2--3 keV), the PD varies between $\sim$2\% and $\sim$18\% and the PA spans roughly between $-16$\degr\ and $70$\degr. 
At higher energies (4--8 keV), the PD ranges between $\sim$3\% and $\sim$12\%, while the PA shows less constrained behavior, spanning the entire possible range between $-90$\degr and $90$\degr.

\item We are able to determine the geometry of the pulsar by modeling the pulse-phase dependence of the PA using the RVM separately for the low and high energy ranges. For the 2--3 keV energy range, we constrained the pulsar geometry in \fourU. 
In particular, the model favors a high $>$50\degr\ spin-axis inclination, which is consistent with alignment to the binary orbital plane that is also has a high inclination of $67\degr$.
The 4--8 keV energy range only provides weak constraints on the pulsar geometry due to lower statistics. The results of the RVM fitting similarly demonstrate the complex nature of energy-dependent polarization properties of \fourU.
\end{enumerate}

\begin{acknowledgements}
The Imaging X-ray Polarimetry Explorer (IXPE) is a joint US and Italian mission. 
This research used data products provided by the IXPE Team and distributed with additional software tools by the High-Energy Astrophysics Science Archive Research Center (HEASARC), at NASA Goddard Space Flight Center.
This research has been supported by the Vilho, Yrjö, and Kalle Väisälä foundation (SVF), the Ministry of Science and Higher Education grant 075-15-2024-647 (SST, JP), the EDUFI Fellowship and Jenny and Antti Wihuri Foundation (AS), and the UKRI Stephen Hawking fellowship (AAM). 
ADM and FLM are supported by the Italian Space Agency (Agenzia Spaziale Italiana, ASI) through contract ASI-INAF-2022-19-HH.0, by the Istituto Nazionale di Astrofisica (INAF) in Italy, and by MAECI with grant CN24GR08 “GRBAXP: Guangxi-Rome Bilateral Agreement for X-ray Polarimetry in Astrophysics”. 
RMEK acknowledges funding from UKRI Science and Technology Facilities Council (STFC) for a PhD studentship (ST/W507891/1), an STFC Long Term Attachment fund, MIT and Dr Herman L. Marshall for travel bursary support (supported in part by MIT Corporate Relations and its Industrial Liaison Program. 
MN is a Fonds de Recherche du Quebec -- Nature et Technologies (FRQNT) postdoctoral fellow. 
VFS thanks  Deutsche  Forschungsgemeinschaft (DFG) grant WE 1312/59-1  for support. 
\end{acknowledgements}

\bibliographystyle{yahapj}
\bibliography{refs}

\begin{thebibliography}{}
\providecommand\natexlab[1]{#1}
\providecommand\JournalTitle[1]{#1}

\bibitem[{{Ahlberg} {et~al.}(2024){Ahlberg}, {Kravtsov}, \& {Poutanen}}]{Ahlberg24}
{Ahlberg}, V., {Kravtsov}, V., \& {Poutanen}, J. 2024, \href{http://dx.doi.org/10.1051/0004-6361/202450131}{\JournalTitle{\aap}, 688, A220}

\bibitem[{{Arnaud}(1996)}]{xspec1996}
{Arnaud}, K.~A. 1996, in ASP Conf. Ser., Vol. 101, Astronomical Data Analysis Software and Systems V, ed. G.~H. {Jacoby} \& J.~{Barnes} (San Francisco: Astron. Soc. Pac.), 17

\bibitem[{{Bailer-Jones} {et~al.}(2021){Bailer-Jones}, {Rybizki}, {Fouesneau}, {Demleitner}, \& {Andrae}}]{Gaia-phot}
{Bailer-Jones}, C.~A.~L., {Rybizki}, J., {Fouesneau}, M., {Demleitner}, M., \& {Andrae}, R. 2021, \href{http://dx.doi.org/10.3847/1538-3881/abd806}{\JournalTitle{\aj}, 161, 147}

\bibitem[{{Baldini} {et~al.}(2021){Baldini}, {Barbanera}, {Bellazzini}, {Bonino}, {Borotto}, {Brez}, {Caporale}, {Cardelli}, {Castellano}, {Ceccanti}, {Citraro}, {Di Lalla}, {Latronico}, {Lucchesi}, {Magazz{\`u}}, {Magazz{\`u}}, {Maldera}, {Manfreda}, {Marengo}, {Marrocchesi}, {Mereu}, {Minuti}, {Mosti}, {Nasimi}, {Nuti}, {Oppedisano}, {Orsini}, {Pesce-Rollins}, {Pinchera}, {Profeti}, {Sgr{\`o}}, {Spandre}, {Tardiola}, {Zanetti}, {Amici}, {Andersson}, {Attin{\`a}}, {Bachetti}, {Baumgartner}, {Brienza}, {Carpentiero}, {Castronuovo}, {Cavalli}, {Cavazzuti}, {Centrone}, {Costa}, {D'Alba}, {D'Amico}, {Del Monte}, {Di Cosimo}, {Di Marco}, {Di Persio}, {Donnarumma}, {Evangelista}, {Fabiani}, {Ferrazzoli}, {Kitaguchi}, {La Monaca}, {Lefevre}, {Loffredo}, {Lorenzi}, {Mangraviti}, {Matt}, {Meilahti}, {Morbidini}, {Muleri}, {Nakano}, {Negri}, {Nenonen}, {O'Dell}, {Perri}, {Piazzolla}, {Pieraccini}, {Pilia}, {Puccetti}, {Ramsey}, {Rankin}, {Ratheesh}, {Rubini}, {Santoli}, {Sarra}, {Scalise}, {Sciortino}, {Soffitta},
  {Tamagawa}, {Tennant}, {Tobia}, {Trois}, {Uchiyama}, {Vimercati}, {Weisskopf}, {Xie}, {Zanetti}, \& {Zhou}}]{2021APh...13302628B}
{Baldini}, L., {Barbanera}, M., {Bellazzini}, R., {et~al.} 2021, \href{http://dx.doi.org/10.1016/j.astropartphys.2021.102628}{\JournalTitle{Astroparticle Physics}, 133, 102628}

\bibitem[{{Baykal} {et~al.}(2006){Baykal}, {Inam}, \& {Beklen}}]{2006A&A...453.1037B}
{Baykal}, A., {Inam}, S.~{\c{C}}., \& {Beklen}, E. 2006, \href{http://dx.doi.org/10.1051/0004-6361:20054616}{\JournalTitle{\aap}, 453, 1037}

\bibitem[{{Becker} {et~al.}(1977){Becker}, {Swank}, {Boldt}, {Holt}, {Pravdo}, {Saba}, \& {Serlemitsos}}]{1977ApJ...216L..11B}
{Becker}, R.~H., {Swank}, J.~H., {Boldt}, E.~A., {et~al.} 1977, \href{http://dx.doi.org/10.1086/182498}{\JournalTitle{\apjl}, 216, L11}

\bibitem[{{Bulik} {et~al.}(1992){Bulik}, {Meszaros}, {Woo}, {Hagase}, \& {Makishima}}]{Bulik1992}
{Bulik}, T., {Meszaros}, P., {Woo}, J.~W., {Hagase}, F., \& {Makishima}, K. 1992, \href{http://dx.doi.org/10.1086/171676}{\JournalTitle{\apj}, 395, 564}

\bibitem[{{Bulik} {et~al.}(1995){Bulik}, {Riffert}, {Meszaros}, {Makishima}, {Mihara}, \& {Thomas}}]{Bulik1995}
{Bulik}, T., {Riffert}, H., {Meszaros}, P., {et~al.} 1995, \href{http://dx.doi.org/10.1086/175614}{\JournalTitle{\apj}, 444, 405}

\bibitem[{{Caiazzo} \& {Heyl}(2021)}]{2021MNRAS.501..109C}
{Caiazzo}, I., \& {Heyl}, J. 2021, \href{http://dx.doi.org/10.1093/mnras/staa3428}{\JournalTitle{\mnras}, 501, 109}

\bibitem[{{Clark}(2000)}]{2000ApJ...542L.131C}
{Clark}, G.~W. 2000, \href{http://dx.doi.org/10.1086/312926}{\JournalTitle{\apjl}, 542, L131}

\bibitem[{{Clark} {et~al.}(1990){Clark}, {Woo}, {Nagase}, {Makishima}, \& {Sakao}}]{1990ApJ...353..274C}
{Clark}, G.~W., {Woo}, J.~W., {Nagase}, F., {Makishima}, K., \& {Sakao}, T. 1990, \href{http://dx.doi.org/10.1086/168614}{\JournalTitle{\apj}, 353, 274}

\bibitem[{{Davison}(1977)}]{1977MNRAS.179P..35D}
{Davison}, P.~J.~N. 1977, \href{http://dx.doi.org/10.1093/mnras/179.1.35P}{\JournalTitle{\mnras}, 179, 35P}

\bibitem[{{Di Marco} {et~al.}(2022{\natexlab{a}}){Di Marco}, {Costa}, {Muleri}, {Soffitta}, {Fabiani}, {La Monaca}, {Rankin}, {Xie}, {Bachetti}, {Baldini}, {Baumgartner}, {Bellazzini}, {Brez}, {Castellano}, {Del Monte}, {Di Lalla}, {Ferrazzoli}, {Latronico}, {Maldera}, {Manfreda}, {O'Dell}, {Perri}, {Pesce-Rollins}, {Puccetti}, {Ramsey}, {Ratheesh}, {Sgr{\`o}}, {Spandre}, {Tennant}, {Tobia}, {Trois}, \& {Weisskopf}}]{DiMarco2022}
{Di Marco}, A., {Costa}, E., {Muleri}, F., {et~al.} 2022{\natexlab{a}}, \href{http://dx.doi.org/10.3847/1538-3881/ac51c9}{\JournalTitle{\aj}, 163, 170}

\bibitem[{{Di Marco} {et~al.}(2022{\natexlab{b}}){Di Marco}, {Fabiani}, {La Monaca}, {Muleri}, {Rankin}, {Soffitta}, {Xie}, {Amici}, {attin{\`a}}, {Bachetti}, {Baldini}, {Barbanera}, {Baumgartner}, {Bellazzini}, {Borotto}, {Brez}, {Brienza}, {Caporale}, {Cardelli}, {Carpentiero}, {Castellano}, {Castronuovo}, {Cavalli}, {Cavazzuti}, {Ceccanti}, {Centrone}, {Citraro}, {Costa}, {D'Alba}, {D'Amico}, {Del Monte}, {Di Cosimo}, {Di Lalla}, {Di Persio}, {Donnarumma}, {Evangelista}, {Ferrazzoli}, {Latronico}, {Lefevre}, {Loffredo}, {Lorenzi}, {Lucchesi}, {Magazz{\`u}}, {Magazz{\`u}}, {Maldera}, {Manfreda}, {Mangraviti}, {Marengo}, {Matt}, {Mereu}, {Minuti}, {Morbidini}, {Mosti}, {Nasimi}, {Negri}, {Nuti}, {O'Dell}, {Orsini}, {Perri}, {Pesce-Rollins}, {Piazzolla}, {Pieraccini}, {Pilia}, {Pinchera}, {Profeti}, {Puccetti}, {Ramsey}, {Ratheesh}, {Rubini}, {Santoli}, {Sarra}, {Scalise}, {Sciortino}, {Sgr{\`o}}, {Spandre}, {Tardiola}, {Tennant}, {Tobia}, {Trois}, {Vimercati}, {Weisskopf}, {Zanetti}, \&
  {Zanetti}}]{DiMarco22b}
{Di Marco}, A., {Fabiani}, S., {La Monaca}, F., {et~al.} 2022{\natexlab{b}}, \href{http://dx.doi.org/10.3847/1538-3881/ac7719}{\JournalTitle{\aj}, 164, 103}

\bibitem[{{Di Marco} {et~al.}(2023){Di Marco}, {Soffitta}, {Costa}, {Ferrazzoli}, {La Monaca}, {Rankin}, {Ratheesh}, {Xie}, {Baldini}, {Del Monte}, {Ehlert}, {Fabiani}, {Kim}, {Muleri}, {O'Dell}, {Ramsey}, {Rubini}, {Sgr{\`o}}, {Silvestri}, {Tennant}, \& {Weisskopf}}]{DiMarco2023}
{Di Marco}, A., {Soffitta}, P., {Costa}, E., {et~al.} 2023, \href{http://dx.doi.org/10.3847/1538-3881/acba0f}{\JournalTitle{\aj}, 165, 143}

\bibitem[{{Doroshenko} {et~al.}(2022){Doroshenko}, {Poutanen}, {Tsygankov}, {Suleimanov}, {Bachetti}, {Caiazzo}, {Costa}, {Di Marco}, {Heyl}, {La Monaca}, {Muleri}, {Mushtukov}, {Pavlov}, {Ramsey}, {Rankin}, {Santangelo}, {Soffitta}, {Staubert}, {Weisskopf}, {Zane}, {Agudo}, {Antonelli}, {Baldini}, {Baumgartner}, {Bellazzini}, {Bianchi}, {Bongiorno}, {Bonino}, {Brez}, {Bucciantini}, {Capitanio}, {Castellano}, {Cavazzuti}, {Ciprini}, {De Rosa}, {Del Monte}, {Di Gesu}, {Di Lalla}, {Donnarumma}, {Dov{\v{c}}iak}, {Ehlert}, {Enoto}, {Evangelista}, {Fabiani}, {Ferrazzoli}, {Garcia}, {Gunji}, {Hayashida}, {Iwakiri}, {Jorstad}, {Karas}, {Kitaguchi}, {Kolodziejczak}, {Krawczynski}, {Latronico}, {Liodakis}, {Maldera}, {Manfreda}, {Marin}, {Marinucci}, {Marscher}, {Marshall}, {Matt}, {Mitsuishi}, {Mizuno}, {Ng}, {O'Dell}, {Omodei}, {Oppedisano}, {Papitto}, {Peirson}, {Perri}, {Pesce-Rollins}, {Pilia}, {Possenti}, {Puccetti}, {Ratheesh}, {Romani}, {Sgr{\`o}}, {Slane}, {Spandre}, {Sunyaev}, {Tamagawa}, {Tavecchio},
  {Taverna}, {Tawara}, {Tennant}, {Thomas}, {Tombesi}, {Trois}, {Turolla}, {Vink}, {Wu}, \& {Xie}}]{2022NatAs...6.1433D}
{Doroshenko}, V., {Poutanen}, J., {Tsygankov}, S.~S., {et~al.} 2022, \href{http://dx.doi.org/10.1038/s41550-022-01799-5}{\JournalTitle{Nature Astronomy}, 6, 1433}

\bibitem[{{Doroshenko} {et~al.}(2023){Doroshenko}, {Poutanen}, {Heyl}, {Tsygankov}, {Caiazzo}, {Turolla}, {Veledina}, {Weisskopf}, {Forsblom}, {Gonz{\'a}lez-Caniulef}, {Loktev}, {Malacaria}, {Mushtukov}, {Suleimanov}, {Lutovinov}, {Mereminskiy}, {Molkov}, {Salganik}, {Santangelo}, {Berdyugin}, {Kravtsov}, {Nitindala}, {Agudo}, {Antonelli}, {Bachetti}, {Baldini}, {Baumgartner}, {Bellazzini}, {Bianchi}, {Bongiorno}, {Bonino}, {Brez}, {Bucciantini}, {Capitanio}, {Castellano}, {Cavazzuti}, {Chen}, {Ciprini}, {Costa}, {De Rosa}, {Del Monte}, {Di Gesu}, {Di Lalla}, {Di Marco}, {Donnarumma}, {Dov{\v{c}}iak}, {Ehlert}, {Enoto}, {Evangelista}, {Fabiani}, {Ferrazzoli}, {Garc{\'\i}a}, {Gunji}, {Hayashida}, {Iwakiri}, {Jorstad}, {Kaaret}, {Karas}, {Kislat}, {Kitaguchi}, {Kolodziejczak}, {Krawczynski}, {La Monaca}, {Latronico}, {Liodakis}, {Maldera}, {Manfreda}, {Marin}, {Marinucci}, {Marscher}, {Marshall}, {Massaro}, {Matt}, {Mitsuishi}, {Mizuno}, {Muleri}, {Negro}, {Ng}, {O'Dell}, {Omodei}, {Oppedisano}, {Papitto},
  {Pavlov}, {Peirson}, {Perri}, {Pesce-Rollins}, {Petrucci}, {Pilia}, {Possenti}, {Puccetti}, {Ramsey}, {Rankin}, {Ratheesh}, {Roberts}, {Romani}, {Sgr{\`o}}, {Slane}, {Soffitta}, {Spandre}, {Swartz}, {Tamagawa}, {Tavecchio}, {Taverna}, {Tawara}, {Tennant}, {Thomas}, {Tombesi}, {Trois}, {Vink}, {Wu}, {Xie}, \& {Zane}}]{Doroshenko23}
{Doroshenko}, V., {Poutanen}, J., {Heyl}, J., {et~al.} 2023, \href{http://dx.doi.org/10.1051/0004-6361/202347088}{\JournalTitle{\aap}, 677, A57}

\bibitem[{{Falanga} {et~al.}(2015){Falanga}, {Bozzo}, {Lutovinov}, {Bonnet-Bidaud}, {Fetisova}, \& {Puls}}]{2015A&A...577A.130F}
{Falanga}, M., {Bozzo}, E., {Lutovinov}, A., {et~al.} 2015, \href{http://dx.doi.org/10.1051/0004-6361/201425191}{\JournalTitle{\aap}, 577, A130}

\bibitem[{{Foreman-Mackey} {et~al.}(2013){Foreman-Mackey}, {Hogg}, {Lang}, \& {Goodman}}]{2013PASP..125..306F}
{Foreman-Mackey}, D., {Hogg}, D.~W., {Lang}, D., \& {Goodman}, J. 2013, \href{http://dx.doi.org/10.1086/670067}{\JournalTitle{\pasp}, 125, 306}

\bibitem[{{Forsblom} {et~al.}(2025){Forsblom}, {Tsygankov}, {Suleimanov}, {Mushtukov}, \& {Poutanen}}]{Forsblom25}
{Forsblom}, S.~V., {Tsygankov}, S.~S., {Suleimanov}, V.~F., {Mushtukov}, A.~A., \& {Poutanen}, J. 2025, \href{http://dx.doi.org/10.48550/arXiv.2501.14324}{\JournalTitle{arXiv e-prints}, arXiv:2501.14324}

\bibitem[{{Forsblom} {et~al.}(2023){Forsblom}, {Poutanen}, {Tsygankov}, {Bachetti}, {Di Marco}, {Doroshenko}, {Heyl}, {La Monaca}, {Malacaria}, {Marshall}, {Muleri}, {Mushtukov}, {Pilia}, {Rogantini}, {Suleimanov}, {Taverna}, {Xie}, {Agudo}, {Antonelli}, {Baldini}, {Baumgartner}, {Bellazzini}, {Bianchi}, {Bongiorno}, {Bonino}, {Brez}, {Bucciantini}, {Capitanio}, {Castellano}, {Cavazzuti}, {Chen}, {Ciprini}, {Costa}, {De Rosa}, {Del Monte}, {Di Gesu}, {Di Lalla}, {Donnarumma}, {Dov{\v{c}}iak}, {Ehlert}, {Enoto}, {Evangelista}, {Fabiani}, {Ferrazzoli}, {Garcia}, {Gunji}, {Hayashida}, {Iwakiri}, {Jorstad}, {Kaaret}, {Karas}, {Kitaguchi}, {Kolodziejczak}, {Krawczynski}, {Latronico}, {Liodakis}, {Maldera}, {Manfreda}, {Marin}, {Marinucci}, {Marscher}, {Matt}, {Mitsuishi}, {Mizuno}, {Negro}, {Ng}, {O'Dell}, {Omodei}, {Oppedisano}, {Papitto}, {Pavlov}, {Peirson}, {Perri}, {Pesce-Rollins}, {Petrucci}, {Possenti}, {Puccetti}, {Ramsey}, {Rankin}, {Ratheesh}, {Roberts}, {Romani}, {Sgr{\`o}}, {Slane}, {Soffitta},
  {Spandre}, {Sunyaev}, {Swartz}, {Tamagawa}, {Tavecchio}, {Tawara}, {Tennant}, {Thomas}, {Tombesi}, {Trois}, {Turolla}, {Vink}, {Weisskopf}, {Wu}, {Zane}, \& {IXPE Collaboration}}]{Forsblom23}
{Forsblom}, S.~V., {Poutanen}, J., {Tsygankov}, S.~S., {et~al.} 2023, \href{http://dx.doi.org/10.3847/2041-8213/acc391}{\JournalTitle{\apjl}, 947, L20}

\bibitem[{{Forsblom} {et~al.}(2024){Forsblom}, {Tsygankov}, {Poutanen}, {Doroshenko}, {Mushtukov}, {Ng}, {Ravi}, {Marshall}, {Di Marco}, {La Monaca}, {Malacaria}, {Mastroserio}, {Loktev}, {Possenti}, {Suleimanov}, {Taverna}, {Agudo}, {Antonelli}, {Bachetti}, {Baldini}, {Baumgartner}, {Bellazzini}, {Bianchi}, {Bongiorno}, {Bonino}, {Brez}, {Bucciantini}, {Capitanio}, {Castellano}, {Cavazzuti}, {Chen}, {Ciprini}, {Costa}, {De Rosa}, {Del Monte}, {Di Gesu}, {Di Lalla}, {Donnarumma}, {Dovciak}, {Ehlert}, {Enoto}, {Evangelista}, {Fabiani}, {Ferrazzoli}, {Garcia}, {Gunji}, {Hayashida}, {Heyl}, {Iwakiri}, {Jorstad}, {Kaaret}, {Karas}, {Kislat}, {Kitaguchi}, {Kolodziejczak}, {Krawczynski}, {Latronico}, {Liodakis}, {Maldera}, {Manfreda}, {Marin}, {Marinucci}, {Marscher}, {Massaro}, {Matt}, {Mitsuishi}, {Mizuno}, {Muleri}, {Negro}, {Ng}, {O'Dell}, {Omodei}, {Oppedisano}, {Papitto}, {Pavlov}, {Peirson}, {Perri}, {Pesce-Rollins}, {Petrucci}, {Pilia}, {Puccetti}, {Ramsey}, {Rankin}, {Ratheesh}, {Roberts}, {Romani},
  {Sgro}, {Slane}, {Soffitta}, {Spandre}, {Swartz}, {Tamagawa}, {Tavecchio}, {Tawara}, {Tennant}, {Thomas}, {Tombesi}, {Trois}, {Turolla}, {Vink}, {Weisskopf}, {Wu}, {Xie}, \& {Zane}}]{Forsblom24}
{Forsblom}, S.~V., {Tsygankov}, S.~S., {Poutanen}, J., {et~al.} 2024, \href{http://dx.doi.org/10.1051/0004-6361/202450937}{\JournalTitle{\aap}, 691, A216}

\bibitem[{{Harding} \& {Lai}(2006)}]{2006RPPh...69.2631H}
{Harding}, A.~K., \& {Lai}, D. 2006, \href{http://dx.doi.org/10.1088/0034-4885/69/9/R03}{\JournalTitle{Reports on Progress in Physics}, 69, 2631}

\bibitem[{{Hemphill} {et~al.}(2019){Hemphill}, {Rothschild}, {Cheatham}, {F{\"u}rst}, {Kretschmar}, {K{\"u}hnel}, {Pottschmidt}, {Staubert}, {Wilms}, \& {Wolff}}]{2019ApJ...873...62H}
{Hemphill}, P.~B., {Rothschild}, R.~E., {Cheatham}, D.~M., {et~al.} 2019, \href{http://dx.doi.org/10.3847/1538-4357/ab03d3}{\JournalTitle{\apj}, 873, 62}

\bibitem[{{Heyl} \& {Caiazzo}(2018)}]{Heyl2018}
{Heyl}, J., \& {Caiazzo}, I. 2018, \href{http://dx.doi.org/10.3390/galaxies6030076}{\JournalTitle{Galaxies}, 6, 76}

\bibitem[{{Heyl} {et~al.}(2024){Heyl}, {Doroshenko}, {Gonz{\'a}lez-Caniulef}, {Caiazzo}, {Poutanen}, {Mushtukov}, {Tsygankov}, {Kirmizibayrak}, {Bachetti}, {Pavlov}, {Forsblom}, {Malacaria}, {Suleimanov}, {Agudo}, {Antonelli}, {Baldini}, {Baumgartner}, {Bellazzini}, {Bianchi}, {Bongiorno}, {Bonino}, {Brez}, {Bucciantini}, {Capitanio}, {Castellano}, {Cavazzuti}, {Chen}, {Ciprini}, {Costa}, {De Rosa}, {Del Monte}, {Di Gesu}, {Di Lalla}, {Di Marco}, {Donnarumma}, {Dov{\v{c}}iak}, {Ehlert}, {Enoto}, {Evangelista}, {Fabiani}, {Ferrazzoli}, {Garcia}, {Gunji}, {Hayashida}, {Iwakiri}, {Jorstad}, {Kaaret}, {Karas}, {Kislat}, {Kitaguchi}, {Kolodziejczak}, {Krawczynski}, {La Monaca}, {Latronico}, {Liodakis}, {Maldera}, {Manfreda}, {Marin}, {Marinucci}, {Marscher}, {Marshall}, {Massaro}, {Matt}, {Mitsuishi}, {Mizuno}, {Muleri}, {Negro}, {Ng}, {O'Dell}, {Omodei}, {Oppedisano}, {Papitto}, {Peirson}, {Perri}, {Pesce-Rollins}, {Petrucci}, {Pilia}, {Possenti}, {Puccetti}, {Ramsey}, {Rankin}, {Ratheesh}, {Roberts}, {Romani},
  {Sgr{\`o}}, {Slane}, {Soffitta}, {Spandre}, {Swartz}, {Tamagawa}, {Tavecchio}, {Taverna}, {Tawara}, {Tennant}, {Thomas}, {Tombesi}, {Trois}, {Turolla}, {Vink}, {Weisskopf}, {Wu}, {Xie}, \& {Zane}}]{Heyl2024}
{Heyl}, J., {Doroshenko}, V., {Gonz{\'a}lez-Caniulef}, D., {et~al.} 2024, \href{http://dx.doi.org/10.1038/s41550-024-02295-8}{\JournalTitle{Nature Astronomy}, 8, 1047}

\bibitem[{{Heyl} \& {Shaviv}(2002)}]{Heyl2002}
{Heyl}, J.~S., \& {Shaviv}, N.~J. 2002, \href{http://dx.doi.org/10.1103/PhysRevD.66.023002}{\JournalTitle{\prd}, 66, 023002}

\bibitem[{{Kislat} {et~al.}(2015){Kislat}, {Clark}, {Beilicke}, \& {Krawczynski}}]{2015-Kislat}
{Kislat}, F., {Clark}, B., {Beilicke}, M., \& {Krawczynski}, H. 2015, \href{http://dx.doi.org/10.1016/j.astropartphys.2015.02.007}{\JournalTitle{Astroparticle Physics}, 68, 45}

\bibitem[{{Malacaria} {et~al.}(2020){Malacaria}, {Jenke}, {Roberts}, {Wilson-Hodge}, {Cleveland}, {Mailyan}, \& {GBM Accreting Pulsars Program Team}}]{2020ApJ...896...90M}
{Malacaria}, C., {Jenke}, P., {Roberts}, O.~J., {et~al.} 2020, \href{http://dx.doi.org/10.3847/1538-4357/ab855c}{\JournalTitle{\apj}, 896, 90}

\bibitem[{{Malacaria} {et~al.}(2023){Malacaria}, {Heyl}, {Doroshenko}, {Tsygankov}, {Poutanen}, {Forsblom}, {Capitanio}, {Di Marco}, {Du}, {Ducci}, {La Monaca}, {Lutovinov}, {Marshall}, {Mereminskiy}, {Molkov}, {Mushtukov}, {Ng}, {Petrucci}, {Santangelo}, {Shtykovsky}, {Suleimanov}, {Agudo}, {Antonelli}, {Bachetti}, {Baldini}, {Baumgartner}, {Bellazzini}, {Bianchi}, {Bongiorno}, {Bonino}, {Brez}, {Bucciantini}, {Castellano}, {Cavazzuti}, {Chen}, {Ciprini}, {Costa}, {De Rosa}, {Del Monte}, {Di Gesu}, {Di Lalla}, {Donnarumma}, {Dov{\v{c}}iak}, {Ehlert}, {Enoto}, {Evangelista}, {Fabiani}, {Ferrazzoli}, {Garcia}, {Gunji}, {Hayashida}, {Iwakiri}, {Jorstad}, {Kaaret}, {Karas}, {Kislat}, {Kitaguchi}, {Kolodziejczak}, {Krawczynski}, {Latronico}, {Liodakis}, {Maldera}, {Manfreda}, {Marin}, {Marinucci}, {Marscher}, {Massaro}, {Matt}, {Mitsuishi}, {Mizuno}, {Muleri}, {Negro}, {Ng}, {O'Dell}, {Omodei}, {Oppedisano}, {Papitto}, {Pavlov}, {Peirson}, {Perri}, {Pesce-Rollins}, {Pilia}, {Possenti}, {Puccetti}, {Ramsey},
  {Rankin}, {Ratheesh}, {Roberts}, {Romani}, {Sgr{\`o}}, {Slane}, {Soffitta}, {Spandre}, {Swartz}, {Tamagawa}, {Tavecchio}, {Taverna}, {Tawara}, {Tennant}, {Thomas}, {Tombesi}, {Trois}, {Turolla}, {Vink}, {Weisskopf}, {Wu}, {Xie}, \& {Zane}}]{2023A&A...675A..29M}
{Malacaria}, C., {Heyl}, J., {Doroshenko}, V., {et~al.} 2023, \href{http://dx.doi.org/10.1051/0004-6361/202346581}{\JournalTitle{\aap}, 675, A29}

\bibitem[{{Marshall} {et~al.}(2022){Marshall}, {Ng}, {Rogantini}, {Heyl}, {Tsygankov}, {Poutanen}, {Costa}, {Zane}, {Malacaria}, {Agudo}, {Antonelli}, {Bachetti}, {Baldini}, {Baumgartner}, {Bellazzini}, {Bianchi}, {Bongiorno}, {Bonino}, {Brez}, {Bucciantini}, {Capitanio}, {Castellano}, {Cavazzuti}, {Ciprini}, {De Rosa}, {Del Monte}, {Di Gesu}, {Di Lalla}, {Di Marco}, {Donnarumma}, {Doroshenko}, {Dov{\v{c}}iak}, {Ehlert}, {Enoto}, {Evangelista}, {Fabiani}, {Ferrazzoli}, {Garcia}, {Gunji}, {Hayashida}, {Iwakiri}, {Jorstad}, {Karas}, {Kitaguchi}, {Kolodziejczak}, {Krawczynski}, {La Monaca}, {Latronico}, {Liodakis}, {Maldera}, {Manfreda}, {Marin}, {Marinucci}, {Marscher}, {Matt}, {Mitsuishi}, {Mizuno}, {Muleri}, {Ng}, {O'Dell}, {Omodei}, {Oppedisano}, {Papitto}, {Pavlov}, {Peirson}, {Perri}, {Pesce-Rollins}, {Petrucci}, {Pilia}, {Possenti}, {Puccetti}, {Ramsey}, {Rankin}, {Ratheesh}, {Romani}, {Sgr{\`o}}, {Slane}, {Soffitta}, {Spandre}, {Tamagawa}, {Tavecchio}, {Taverna}, {Tawara}, {Tennant}, {Thomas}, {Tombesi},
  {Trois}, {Turolla}, {Vink}, {Weisskopf}, {Wu}, {Xie}, {IXPE Collaboration}, {Schulz}, \& {Chakrabarty}}]{2022ApJ...940...70M}
{Marshall}, H.~L., {Ng}, M., {Rogantini}, D., {et~al.} 2022, \href{http://dx.doi.org/10.3847/1538-4357/ac98c2}{\JournalTitle{\apj}, 940, 70}

\bibitem[{{Matsuoka} {et~al.}(2009){Matsuoka}, {Kawasaki}, {Ueno}, {Tomida}, {Kohama}, {Suzuki}, {Adachi}, {Ishikawa}, {Mihara}, {Sugizaki}, {Isobe}, {Nakagawa}, {Tsunemi}, {Miyata}, {Kawai}, {Kataoka}, {Morii}, {Yoshida}, {Negoro}, {Nakajima}, {Ueda}, {Chujo}, {Yamaoka}, {Yamazaki}, {Nakahira}, {You}, {Ishiwata}, {Miyoshi}, {Eguchi}, {Hiroi}, {Katayama}, \& {Ebisawa}}]{Matsuoka2009}
{Matsuoka}, M., {Kawasaki}, K., {Ueno}, S., {et~al.} 2009, \href{http://dx.doi.org/10.1093/pasj/61.5.999}{\JournalTitle{\pasj}, 61, 999}

\bibitem[{{Matt}(1993)}]{Matt93}
{Matt}, G. 1993, \href{http://dx.doi.org/10.1093/mnras/260.3.663}{\JournalTitle{\mnras}, 260, 663}

\bibitem[{{Meszaros} {et~al.}(1988){Meszaros}, {Novick}, {Szentgyorgyi}, {Chanan}, \& {Weisskopf}}]{1988ApJ...324.1056M}
{Meszaros}, P., {Novick}, R., {Szentgyorgyi}, A., {Chanan}, G.~A., \& {Weisskopf}, M.~C. 1988, \href{http://dx.doi.org/10.1086/165962}{\JournalTitle{\apj}, 324, 1056}

\bibitem[{{Mushtukov} \& {Tsygankov}(2024)}]{MushtukovTsygankov2024}
{Mushtukov}, A., \& {Tsygankov}, S. 2024, \href{http://dx.doi.org/10.1007/978-981-19-6960-7_104}{in Handbook of X-ray and Gamma-ray Astrophysics, ed. C.~{Bambi} \& A.~{Santangelo}} (Singapore: Springer), 4105

\bibitem[{{Mushtukov} {et~al.}(2023){Mushtukov}, {Tsygankov}, {Poutanen}, {Doroshenko}, {Salganik}, {Costa}, {Di Marco}, {Heyl}, {La Monaca}, {Lutovinov}, {Mereminsky}, {Papitto}, {Semena}, {Shtykovsky}, {Suleimanov}, {Forsblom}, {Gonz{\'a}lez-Caniulef}, {Malacaria}, {Sunyaev}, {Agudo}, {Antonelli}, {Bachetti}, {Baldini}, {Baumgartner}, {Bellazzini}, {Bianchi}, {Bongiorno}, {Bonino}, {Brez}, {Bucciantini}, {Capitanio}, {Castellano}, {Cavazzuti}, {Chen}, {Ciprini}, {De Rosa}, {Del Monte}, {Di Gesu}, {Di Lalla}, {Donnarumma}, {Dov{\v{c}}iak}, {Ehlert}, {Enoto}, {Evangelista}, {Fabiani}, {Ferrazzoli}, {Garcia}, {Gunji}, {Hayashida}, {Iwakiri}, {Jorstad}, {Kaaret}, {Karas}, {Kislat}, {Kitaguchi}, {Kolodziejczak}, {Krawczynski}, {Latronico}, {Liodakis}, {Maldera}, {Manfreda}, {Marin}, {Marscher}, {Marshall}, {Massaro}, {Matt}, {Mitsuishi}, {Mizuno}, {Muleri}, {Negro}, {Ng}, {O'Dell}, {Omodei}, {Oppedisano}, {Pavlov}, {Peirson}, {Perri}, {Pesce-Rollins}, {Petrucci}, {Pilia}, {Possenti}, {Puccetti}, {Ramsey},
  {Rankin}, {Ratheesh}, {Roberts}, {Romani}, {Sgr{\`o}}, {Slane}, {Soffitta}, {Spandre}, {Swartz}, {Tamagawa}, {Tavecchio}, {Taverna}, {Tawara}, {Tennant}, {Thomas}, {Tombesi}, {Trois}, {Turolla}, {Vink}, {Weisskopf}, {Wu}, {Xie}, \& {Zane}}]{2023MNRAS.524.2004M}
{Mushtukov}, A.~A., {Tsygankov}, S.~S., {Poutanen}, J., {et~al.} 2023, \href{http://dx.doi.org/10.1093/mnras/stad1961}{\JournalTitle{\mnras}, 524, 2004}

\bibitem[{{Naghizadeh-Khouei} \& {Clarke}(1993)}]{1993A&A...274..968N}
{Naghizadeh-Khouei}, J., \& {Clarke}, D. 1993, \JournalTitle{\aap}, 274, 968

\bibitem[{{Neumann} {et~al.}(2023){Neumann}, {Avakyan}, {Doroshenko}, \& {Santangelo}}]{2023-HMXBcat}
{Neumann}, M., {Avakyan}, A., {Doroshenko}, V., \& {Santangelo}, A. 2023, \href{http://dx.doi.org/10.1051/0004-6361/202245728}{\JournalTitle{\aap}, 677, A134}

\bibitem[{{Nitindala} {et~al.}(2025){Nitindala}, {Veledina}, \& {Poutanen}}]{Nitindala25}
{Nitindala}, A.~P., {Veledina}, A., \& {Poutanen}, J. 2025, \href{http://dx.doi.org/10.1051/0004-6361/202453188}{\JournalTitle{\aap}, A230}

\bibitem[{{Poutanen}(2020)}]{2020A&A...641A.166P}
{Poutanen}, J. 2020, \href{http://dx.doi.org/10.1051/0004-6361/202038689}{\JournalTitle{\aap}, 641, A166}

\bibitem[{{Poutanen} {et~al.}(1996){Poutanen}, {Nagendra}, \& {Svensson}}]{Poutanen96}
{Poutanen}, J., {Nagendra}, K.~N., \& {Svensson}, R. 1996, \href{http://dx.doi.org/10.1093/mnras/283.3.892}{\JournalTitle{\mnras}, 283, 892}

\bibitem[{{Poutanen} {et~al.}(2024{\natexlab{a}}){Poutanen}, {Tsygankov}, \& {Forsblom}}]{Poutanen2024Galax}
{Poutanen}, J., {Tsygankov}, S.~S., \& {Forsblom}, S.~V. 2024{\natexlab{a}}, \href{http://dx.doi.org/10.3390/galaxies12040046}{\JournalTitle{Galaxies}, 12, 46}

\bibitem[{{Poutanen} {et~al.}(2024{\natexlab{b}}){Poutanen}, {Tsygankov}, {Doroshenko}, {Forsblom}, {Jenke}, {Kaaret}, {Berdyugin}, {Blinov}, {Kravtsov}, {Liodakis}, {Tzouvanou}, {Di Marco}, {Heyl}, {La Monaca}, {Mushtukov}, {Pavlov}, {Salganik}, {Veledina}, {Weisskopf}, {Zane}, {Loktev}, {Suleimanov}, {Wilson-Hodge}, {Berdyugina}, {Kagitani}, {Piirola}, {Sakanoi}, {Agudo}, {Antonelli}, {Bachetti}, {Baldini}, {Baumgartner}, {Bellazzini}, {Bianchi}, {Bongiorno}, {Bonino}, {Brez}, {Bucciantini}, {Capitanio}, {Castellano}, {Cavazzuti}, {Chen}, {Ciprini}, {Costa}, {De Rosa}, {Del Monte}, {Di Gesu}, {Di Lalla}, {Donnarumma}, {Dovciak}, {Ehlert}, {Enoto}, {Evangelista}, {Fabiani}, {Ferrazzoli}, {Garcia}, {Gunji}, {Hayashida}, {Iwakiri}, {Jorstad}, {Karas}, {Kislat}, {Kitaguchi}, {Kolodziejczak}, {Latronico}, {Maldera}, {Manfreda}, {Marin}, {Marinucci}, {Marscher}, {Marshall}, {Massaro}, {Matt}, {Mitsuishi}, {Mizuno}, {Muleri}, {Negro}, {Ng}, {O'Dell}, {Omodei}, {Oppedisano}, {Papitto}, {Peirson}, {Perri},
  {Pesce-Rollins}, {Petrucci}, {Pilia}, {Possenti}, {Puccetti}, {Ramsey}, {Rankin}, {Ratheesh}, {Roberts}, {Romani}, {Sgro}, {Slane}, {Soffitta}, {Spandre}, {Swartz}, {Tamagawa}, {Tavecchio}, {Taverna}, {Tawara}, {Tennant}, {Thomas}, {Tombesi}, {Trois}, {Turolla}, {Vink}, {Wu}, \& {Xie}}]{Poutanen2024}
{Poutanen}, J., {Tsygankov}, S.~S., {Doroshenko}, V., {et~al.} 2024{\natexlab{b}}, \href{http://dx.doi.org/10.1051/0004-6361/202450696}{\JournalTitle{\aap}, 691, A123}

\bibitem[{{Radhakrishnan} \& {Cooke}(1969)}]{1969ApL.....3..225R}
{Radhakrishnan}, V., \& {Cooke}, D.~J. 1969, \JournalTitle{\aplett}, 3, 225

\bibitem[{{Rodes-Roca} {et~al.}(2011){Rodes-Roca}, {Page}, {Torrej{\'o}n}, {Osborne}, \& {Bernab{\'e}u}}]{2011A&A...526A..64R}
{Rodes-Roca}, J.~J., {Page}, K.~L., {Torrej{\'o}n}, J.~M., {Osborne}, J.~P., \& {Bernab{\'e}u}, G. 2011, \href{http://dx.doi.org/10.1051/0004-6361/201014324}{\JournalTitle{\aap}, 526, A64}

\bibitem[{{Rubin} {et~al.}(1997){Rubin}, {Finger}, {Scott}, \& {Wilson}}]{1997ApJ...488..413R}
{Rubin}, B.~C., {Finger}, M.~H., {Scott}, D.~M., \& {Wilson}, R.~B. 1997, \href{http://dx.doi.org/10.1086/304679}{\JournalTitle{\apj}, 488, 413}

\bibitem[{{Soffitta} {et~al.}(2021){Soffitta}, {Baldini}, {Bellazzini}, {Costa}, {Latronico}, {Muleri}, {Del Monte}, {Fabiani}, {Minuti}, {Pinchera}, {Sgro'}, {Spandre}, {Trois}, {Amici}, {Andersson}, {Attina'}, {Bachetti}, {Barbanera}, {Borotto}, {Brez}, {Brienza}, {Caporale}, {Cardelli}, {Carpentiero}, {Castellano}, {Castronuovo}, {Cavalli}, {Cavazzuti}, {Ceccanti}, {Centrone}, {Ciprini}, {Citraro}, {D'Amico}, {D'Alba}, {Di Cosimo}, {Di Lalla}, {Di Marco}, {Di Persio}, {Donnarumma}, {Evangelista}, {Ferrazzoli}, {Hayato}, {Kitaguchi}, {La Monaca}, {Lefevre}, {Loffredo}, {Lorenzi}, {Lucchesi}, {Magazzu}, {Maldera}, {Manfreda}, {Mangraviti}, {Marengo}, {Matt}, {Mereu}, {Morbidini}, {Mosti}, {Nakano}, {Nasimi}, {Negri}, {Nenonen}, {Nuti}, {Orsini}, {Perri}, {Pesce-Rollins}, {Piazzolla}, {Pilia}, {Profeti}, {Puccetti}, {Rankin}, {Ratheesh}, {Rubini}, {Santoli}, {Sarra}, {Scalise}, {Sciortino}, {Tamagawa}, {Tardiola}, {Tobia}, {Vimercati}, \& {Xie}}]{2021AJ....162..208S}
{Soffitta}, P., {Baldini}, L., {Bellazzini}, R., {et~al.} 2021, \href{http://dx.doi.org/10.3847/1538-3881/ac19b0}{\JournalTitle{\aj}, 162, 208}

\bibitem[{{Suleimanov} {et~al.}(2023){Suleimanov}, {Forsblom}, {Tsygankov}, {Poutanen}, {Doroshenko}, {Doroshenko}, {Capitanio}, {Di Marco}, {Gonz{\'a}lez-Caniulef}, {Heyl}, {La Monaca}, {Lutovinov}, {Molkov}, {Malacaria}, {Mushtukov}, {Shtykovsky}, {Agudo}, {Antonelli}, {Bachetti}, {Baldini}, {Baumgartner}, {Bellazzini}, {Bianchi}, {Bongiorno}, {Bonino}, {Brez}, {Bucciantini}, {Castellano}, {Cavazzuti}, {Chen}, {Ciprini}, {Costa}, {De Rosa}, {Del Monte}, {Di Gesu}, {Di Lalla}, {Donnarumma}, {Dov{\v{c}}iak}, {Ehlert}, {Enoto}, {Evangelista}, {Fabiani}, {Ferrazzoli}, {Garcia}, {Gunji}, {Hayashida}, {Iwakiri}, {Jorstad}, {Kaaret}, {Karas}, {Kislat}, {Kitaguchi}, {Kolodziejczak}, {Krawczynski}, {Latronico}, {Liodakis}, {Maldera}, {Manfreda}, {Marin}, {Marinucci}, {Marscher}, {Marshall}, {Massaro}, {Matt}, {Mitsuishi}, {Mizuno}, {Muleri}, {Negro}, {Ng}, {O'Dell}, {Omodei}, {Oppedisano}, {Papitto}, {Pavlov}, {Peirson}, {Perri}, {Pesce-Rollins}, {Petrucci}, {Pilia}, {Possenti}, {Puccetti}, {Ramsey}, {Rankin},
  {Ratheesh}, {Roberts}, {Romani}, {Sgr{\`o}}, {Slane}, {Soffitta}, {Spandre}, {Swartz}, {Tamagawa}, {Tavecchio}, {Taverna}, {Tawara}, {Tennant}, {Thomas}, {Tombesi}, {Trois}, {Turolla}, {Vink}, {Weisskopf}, {Wu}, {Xie}, \& {Zane}}]{2023A&A...678A.119S}
{Suleimanov}, V.~F., {Forsblom}, S.~V., {Tsygankov}, S.~S., {et~al.} 2023, \href{http://dx.doi.org/10.1051/0004-6361/202346994}{\JournalTitle{\aap}, 678, A119}

\bibitem[{{Tamang} {et~al.}(2024){Tamang}, {Ghising}, {Tobrej}, {Rai}, \& {Paul}}]{Tamang2024}
{Tamang}, R., {Ghising}, M., {Tobrej}, M., {Rai}, B., \& {Paul}, B.~C. 2024, \href{http://dx.doi.org/10.1093/mnras/stad2907}{\JournalTitle{\mnras}, 527, 3164}

\bibitem[{{Taverna} {et~al.}(2015){Taverna}, {Turolla}, {Gonzalez Caniulef}, {Zane}, {Muleri}, \& {Soffitta}}]{Taverna15}
{Taverna}, R., {Turolla}, R., {Gonzalez Caniulef}, D., {et~al.} 2015, \href{http://dx.doi.org/10.1093/mnras/stv2168}{\JournalTitle{\mnras}, 454, 3254}

\bibitem[{{Tsygankov} {et~al.}(2022){Tsygankov}, {Doroshenko}, {Poutanen}, {Heyl}, {Mushtukov}, {Caiazzo}, {Di Marco}, {Forsblom}, {Gonz{\'a}lez-Caniulef}, {Klawin}, {La Monaca}, {Malacaria}, {Marshall}, {Muleri}, {Ng}, {Suleimanov}, {Sunyaev}, {Turolla}, {Agudo}, {Antonelli}, {Bachetti}, {Baldini}, {Baumgartner}, {Bellazzini}, {Bianchi}, {Bongiorno}, {Bonino}, {Brez}, {Bucciantini}, {Capitanio}, {Castellano}, {Cavazzuti}, {Ciprini}, {Costa}, {De Rosa}, {Del Monte}, {Di Gesu}, {Di Lalla}, {Donnarumma}, {Dov{\v{c}}iak}, {Ehlert}, {Enoto}, {Evangelista}, {Fabiani}, {Ferrazzoli}, {Garcia}, {Gunji}, {Hayashida}, {Iwakiri}, {Jorstad}, {Karas}, {Kitaguchi}, {Kolodziejczak}, {Krawczynski}, {Latronico}, {Liodakis}, {Maldera}, {Manfreda}, {Marin}, {Marinucci}, {Marscher}, {Matt}, {Mitsuishi}, {Mizuno}, {Ng}, {O'Dell}, {Omodei}, {Oppedisano}, {Papitto}, {Pavlov}, {Peirson}, {Perri}, {Pesce-Rollins}, {Petrucci}, {Pilia}, {Possenti}, {Puccetti}, {Ramsey}, {Rankin}, {Ratheesh}, {Romani}, {Sgr{\`o}}, {Slane}, {Soffitta},
  {Spandre}, {Tamagawa}, {Tavecchio}, {Taverna}, {Tawara}, {Tennant}, {Thomas}, {Tombesi}, {Trois}, {Vink}, {Weisskopf}, {Wu}, {Xie}, {Zane}, \& {IXPE Collaboration}}]{Tsygankov22}
{Tsygankov}, S.~S., {Doroshenko}, V., {Poutanen}, J., {et~al.} 2022, \href{http://dx.doi.org/10.3847/2041-8213/aca486}{\JournalTitle{\apjl}, 941, L14}

\bibitem[{{Tsygankov} {et~al.}(2023){Tsygankov}, {Doroshenko}, {Mushtukov}, {Poutanen}, {Di Marco}, {Heyl}, {La Monaca}, {Forsblom}, {Malacaria}, {Marshall}, {Suleimanov}, {Svoboda}, {Taverna}, {Ursini}, {Agudo}, {Antonelli}, {Bachetti}, {Baldini}, {Baumgartner}, {Bellazzini}, {Bianchi}, {Bongiorno}, {Bonino}, {Brez}, {Bucciantini}, {Capitanio}, {Castellano}, {Cavazzuti}, {Chen}, {Ciprini}, {Costa}, {De Rosa}, {Del Monte}, {Di Gesu}, {Di Lalla}, {Donnarumma}, {Dov{\v{c}}iak}, {Ehlert}, {Enoto}, {Evangelista}, {Fabiani}, {Ferrazzoli}, {Garcia}, {Gunji}, {Hayashida}, {Iwakiri}, {Jorstad}, {Kaaret}, {Karas}, {Kislat}, {Kitaguchi}, {Kolodziejczak}, {Krawczynski}, {Latronico}, {Liodakis}, {Maldera}, {Manfreda}, {Marin}, {Marinucci}, {Marscher}, {Massaro}, {Matt}, {Mitsuishi}, {Mizuno}, {Muleri}, {Negro}, {Ng}, {O'Dell}, {Omodei}, {Oppedisano}, {Papitto}, {Pavlov}, {Peirson}, {Perri}, {Pesce-Rollins}, {Petrucci}, {Pilia}, {Possenti}, {Puccetti}, {Ramsey}, {Rankin}, {Ratheesh}, {Roberts}, {Romani}, {Sgr{\`o}},
  {Slane}, {Soffitta}, {Spandre}, {Swartz}, {Tamagawa}, {Tavecchio}, {Tawara}, {Tennant}, {Thomas}, {Tombesi}, {Trois}, {Turolla}, {Vink}, {Weisskopf}, {Wu}, {Xie}, \& {Zane}}]{2023A&A...675A..48T}
{Tsygankov}, S.~S., {Doroshenko}, V., {Mushtukov}, A.~A., {et~al.} 2023, \href{http://dx.doi.org/10.1051/0004-6361/202346134}{\JournalTitle{\aap}, 675, A48}

\bibitem[{{van Kerkwijk} {et~al.}(1995){van Kerkwijk}, {van Paradijs}, \& {Zuiderwijk}}]{1995A&A...303..497V}
{van Kerkwijk}, M.~H., {van Paradijs}, J., \& {Zuiderwijk}, E.~J. 1995, \href{http://dx.doi.org/10.48550/arXiv.astro-ph/9505071}{\JournalTitle{\aap}, 303, 497}

\bibitem[{{Weisskopf} {et~al.}(2022){Weisskopf}, {Soffitta}, {Baldini}, {Ramsey}, {O'Dell}, {Romani}, {Matt}, {Deininger}, {Baumgartner}, {Bellazzini}, {Costa}, {Kolodziejczak}, {Latronico}, {Marshall}, {Muleri}, {Bongiorno}, {Tennant}, {Bucciantini}, {Dovciak}, {Marin}, {Marscher}, {Poutanen}, {Slane}, {Turolla}, {Kalinowski}, {Di Marco}, {Fabiani}, {Minuti}, {La Monaca}, {Pinchera}, {Rankin}, {Sgro'}, {Trois}, {Xie}, {Alexander}, {Allen}, {Amici}, {Andersen}, {Antonelli}, {Antoniak}, {Attin{\`a}}, {Barbanera}, {Bachetti}, {Baggett}, {Bladt}, {Brez}, {Bonino}, {Boree}, {Borotto}, {Breeding}, {Brienza}, {Bygott}, {Caporale}, {Cardelli}, {Carpentiero}, {Castellano}, {Castronuovo}, {Cavalli}, {Cavazzuti}, {Ceccanti}, {Centrone}, {Citraro}, {D'Amico}, {D'Alba}, {Di Gesu}, {Del Monte}, {Dietz}, {Di Lalla}, {Persio}, {Dolan}, {Donnarumma}, {Evangelista}, {Ferrant}, {Ferrazzoli}, {Ferrie}, {Footdale}, {Forsyth}, {Foster}, {Garelick}, {Gunji}, {Gurnee}, {Head}, {Hibbard}, {Johnson}, {Kelly}, {Kilaru}, {Lefevre},
  {Roy}, {Loffredo}, {Lorenzi}, {Lucchesi}, {Maddox}, {Magazzu}, {Maldera}, {Manfreda}, {Mangraviti}, {Marengo}, {Marrocchesi}, {Massaro}, {Mauger}, {McCracken}, {McEachen}, {Mize}, {Mereu}, {Mitchell}, {Mitsuishi}, {Morbidini}, {Mosti}, {Nasimi}, {Negri}, {Negro}, {Nguyen}, {Nitschke}, {Nuti}, {Onizuka}, {Oppedisano}, {Orsini}, {Osborne}, {Pacheco}, {Paggi}, {Painter}, {Pavelitz}, {Pentz}, {Piazzolla}, {Perri}, {Pesce-Rollins}, {Peterson}, {Pilia}, {Profeti}, {Puccetti}, {Ranganathan}, {Ratheesh}, {Reedy}, {Root}, {Rubini}, {Ruswick}, {Sanchez}, {Sarra}, {Santoli}, {Scalise}, {Sciortino}, {Schroeder}, {Seek}, {Sosdian}, {Spandre}, {Speegle}, {Tamagawa}, {Tardiola}, {Tobia}, {Thomas}, {Valerie}, {Vimercati}, {Walden}, {Weddendorf}, {Wedmore}, {Welch}, {Zanetti}, \& {Zanetti}}]{Weisskopf2022}
{Weisskopf}, M.~C., {Soffitta}, P., {Baldini}, L., {et~al.} 2022, \href{http://dx.doi.org/10.1117/1.JATIS.8.2.026002}{\JournalTitle{JATIS}, 8, 026002}

\bibitem[{{Wilms} {et~al.}(2000){Wilms}, {Allen}, \& {McCray}}]{Wilms2000}
{Wilms}, J., {Allen}, A., \& {McCray}, R. 2000, \href{http://dx.doi.org/10.1086/317016}{\JournalTitle{\apj}, 542, 914}

\end{thebibliography}

\end{document}